\documentclass[aps,pre,twocolumn,superscriptaddress]{revtex4}
\usepackage[dvips]{graphics}
\usepackage{graphicx}
\usepackage{amsfonts}
\usepackage{amssymb}
\usepackage{amsmath}
\usepackage{subfigure}
\usepackage{color}

\def\figw{7.6cm}
\def\rootfig{./}

\begin{document}

\title{Surface Solitons in Three Dimensions}
\author{Q.E. Hoq}
\affiliation{Department of Mathematics, Western New England College, Springfield, MA,
01119, USA}
\author{R. Carretero-Gonz\'alez}
\affiliation{Nonlinear Dynamical Systems Group\thanks{\texttt{URL}: http://nlds.sdsu.edu}%
, Department of Mathematics and Statistics, and Computational Science
Research Center, San Diego State University, San Diego CA, 92182-7720, USA}
\author{P.G.\ Kevrekidis}
\affiliation{Department of Mathematics and Statistics, University of Massachusetts,
Amherst MA 01003-4515, USA}
\author{B.A. Malomed}
\affiliation{Department of Physical Electronics, Faculty of Engineering, Tel Aviv
University, Tel Aviv 69978, Israel}
\author{D.J. Frantzeskakis}
\affiliation{Department of Physics, University of Athens, Panepistimiopolis, Zografos,
Athens 157 84, Greece}
\author{Yu.V. Bludov}
\affiliation{Centro de F\'{\i}sica Te\'orica e Computacional, Universidade de Lisboa,
Complexo Interdisciplinar, Avenida Professor Gama Pinto 2, Lisboa 1649-003,
Portugal}
\author{V.V. Konotop}
\affiliation{Centro de F\'{\i}sica Te\'orica e Computacional, Universidade de Lisboa,
Complexo Interdisciplinar, Avenida Professor Gama Pinto 2, Lisboa 1649-003,
Portugal}
\affiliation{Departamento de F\'{\i}sica, Faculdade de Ci\^encias, Universidade de
Lisboa, Campo Grande, Ed. C8, Piso 6, Lisboa 1749-016, Portugal.}

\date{To appear in {\em Phys. Rev. E}, 2008}.

\begin{abstract}
We study localized modes on the surface of a three-dimensional dynamical
lattice. The stability of these structures on the surface is investigated
and compared to that in the bulk of the lattice. Typically, the surface
makes the stability region larger, an extreme example of that being the
three-site ``horseshoe''-shaped structure, which is always unstable in the
bulk, while at the surface it is stable near the anti-continuum limit. We
also examine effects of the surface on lattice vortices. For the vortex
placed parallel to the surface this increased stability region feature is
also observed, while the vortex cannot exist in a state normal to the
surface. More sophisticated localized dynamical structures, such as
five-site horseshoes and pyramids, are also considered.
\end{abstract}

\maketitle


\section{Introduction}

Surface waves have been a subject of interest in a variety of contexts,
including surface plasmons in conductors \cite{Barnes} and optical solitons
in waveguide arrays \cite{Christo} in physics, surface waves in isotropic
magnetic gels \cite{Bohlius} in chemistry, water waves in the ocean in
geophysical hydrodynamics, and so on. Quite often, features exhibited by
such wave modes have no analog in the corresponding bulk media, which makes
their study especially relevant. In particular, a great deal of interest has
been drawn to nonlinear surface waves in optics. It was shown theoretically
\cite{Makris} and observed experimentally \cite{Suntsov} that discrete
localized nonlinear waves can be supported at the edge of a semi-infinite
array of nonlinear optical waveguide arrays. Such solitary waves were
predicted to exist not only in self-focusing media (as in the
above-mentioned works), but also between uniform and self-defocusing media~%
\cite{Makris,kartprl}, or between self-focusing and self-defocusing media
(e.g. in \cite{debra}). They have been subsequently observed in media with
quadratic \cite{Sivilogou} and photorefractive \cite{rosberg,smirnov}
nonlinearities. In the two-dimensional (2D) geometry, stable topological
solitons have been predicted in a saturable medium \cite{Kartashov}, which
constitute generalizations to lattice vortex solitons predicted in Ref.~\cite%
{Malomed}. Quasi-discrete vortex solitons have been experimentally observed
in a self-focusing bulk photorefractive medium \cite{Neshev}. Theoretical
predictions for a variety of species of discrete 2D surface solitons~\cite%
{Susanto,Makris2,Kartashov2,BluKon,Vicencio}, corner modes~\cite%
{Makris2,BluKon}, as well as surface breathers~\cite{BluKon}, were reported.
Subsequent work has resulted in the experimental observation of 2D surface
solitons, both fundamental ones and multi-pulse states, in photorefractive
media \cite{christo2d}, as well as in asymmetric waveguide arrays written in
fused silica \cite{kart2d}. Recently, surface solitons in more complex
settings, such as chirped optical lattices in 1D and 2D \cite%
{kart_mol1,kart_mol2}, at interfaces between photonic crystals and
metamaterials \cite{shadrivov}, and in the case of nonlocal nonlinearity
\cite{moti1,moti2}, have emerged.

Nearly all these efforts have been aimed at the study of surface solitons in
1D and 2D geometries. The only 3D setting examined thus far assumed a
truncated bundle of fiber-like waveguides, incorporating the temporal
dynamics in longitudinal direction to produce 3D ``surface light bullets" in
Ref.~\cite{dumitru} (the respective 2D surface structures were examined in
Ref.~\cite{dumitru1}).

Our aim in the present work is to extend the analysis to surface solitons in
genuine 3D lattices. Our setup is relevant to a variety of applications
including, e.g., crystals built of microresonators trapping photons \cite%
{earlier1}, polaritons \cite{earlier2}, or Bose-Einstein
condensates in the vicinity of an edge of a strong 3D optical
lattice \cite{morsch,bloch}. In particular, we report results for
discrete solitons at the surface of a 3D lattice, i.e., 3D
localized states that are similar to relevant objects studied in
the 2D setting of Ref.~\cite{Susanto}. Thus, we will study
localized states such as dipoles and ``horseshoes" abutting on a
set of three lattice sites, but also states that are specific to
the 3D lattice. A variety of species of such solitons is examined
below, and their stability on the surface is compared to that in
the bulk. Some localized structures, such as dipoles, may be
placed either normal or parallel to the surface. We demonstrate
that, typically, the enhanced contact with the surface
\emph{increases} the stability region of the structure.
Physically, this conclusion may be understood by the fact that the
surface reduces the local interactions to fewer neighbors,
rendering the system ``more discrete", hence more stable (by
pushing the medium further away from the continuum limit, where
all solitons would be unstable against the collapse). This effect
is remarkable, e.g., for the three-site horseshoes which are never
stable in the bulk, but get stabilized in the presence of the
surface. However, the surface may also have an adverse effect,
inhibiting the existence of a particular mode. The latter trend is
exemplified by discrete vortices, which, if placed parallel to the
surface, feature enhanced stability as compared to the bulk-mode
counterpart, but cannot exist with the orientation perpendicular
to the surface.
Surface-induced effects of a different kind, which are less specific to
discrete systems, are induced by the interaction of a particular localized
mode with its fictitious ``mirror image". In terms of lattice models, the
approach based on the analysis of the interaction of a real mode with its
image was proposed in Ref. \cite{Christodoulides}.

To formulate the model, we introduce unit vectors $\mathbf{e}_{1}=(1,0,0)$, $%
\mathbf{e}_{2}=(0,1,0)$, and $\mathbf{e}_{3}=(0,0,1)$ and define lattice
sites by $\mathbf{n}=\sum_{j=1}^{3}n_{j}\mathbf{e}_{j}$ with integer $n_{j}$%
. We assume that the lattice occupies a semi-infinite space, $n_{3}\geq 1$,
and its dynamics obeys the discrete nonlinear Schr\"{o}dinger (DNLS)
equation in its usual form,
\begin{equation}
i\dot{\phi}_{\mathbf{n}}+\varepsilon \Delta \phi _{\mathbf{n}}+\sigma |\phi
_{\mathbf{n}}|^{2}\phi _{\mathbf{n}}=0.  \label{schr1}
\end{equation}%
Here $\phi _{\mathbf{n}}$ is a complex discrete field, $\varepsilon $ is the
coupling constant, $\dot{\phi}_{\mathbf{n}}$ stands for the time derivative,
the parameter $\sigma =\pm 1$ determines the sign of the nonlinearity
(focusing or defocusing respectively), and $\Delta \phi _{\mathbf{n}}$ is
the 3D discrete Laplacian:
\begin{equation}
\Delta \phi _{\mathbf{n}}\!\!\equiv \!\!\sum_{j=1}^{3}\left( \phi _{\mathbf{n%
}+\mathbf{e}_{j}}+\phi _{\mathbf{n}-\mathbf{e}_{j}}-2s\phi _{\mathbf{n}%
}\right) ,  \label{stat}
\end{equation}%
for $n_{3}\geq 2$, while for $n_{3}=1$ the term with subscript index $%
\mathbf{n}-\mathbf{e}_{3}$ is to be dropped (note that $\mathbf{e}_{3}$ is
the direction normal to the surface).


It is interesting to point out here that an approach towards
understanding the dynamics of  Eq.~(\ref{schr1}) in the vicinity
of the surface can be based on the 
above-mentioned concept of the fictitious mirror image, 
formally extends the range of $n_{3}$ up to $n_{3}=-\infty$, 
supplementing the equation with the anti-symmetry condition,
\begin{equation}
\phi _{n_{1},n_{2},-n_{3}}\equiv \phi _{n_{1},n_{2}n_{3}}.
\end{equation}
Indeed, this condition implies $\phi _{n_{1},n_{2},0}\equiv 0$, 
which is equivalent to the summation restriction in 
Eq.~(\ref{stat}) as defined above.

To confine the analysis to localized solitary wave modes, we impose zero
boundary conditions, $\phi _{\mathbf{n}}\rightarrow 0$ at $%
n_{1,2}\rightarrow \pm \infty $ and $n_{3}\rightarrow \infty $.
Additionally, $s=\pm 1$ in Eq.~(\ref{stat}) ---this parameter is introduced
for convenience (see Sec.~\ref{small-amp}) and can be freely rescaled using
the transformation $\phi \rightarrow \phi \,e^{i\nu t}$ for an appropriate
choice of $\nu $ and time rescaling. Stationary solutions to Eq.~(\ref{schr1}%
) will be sought for as $\phi _{\mathbf{n}}=\exp {(i\Lambda t)}u_{\mathbf{n}%
} $, where $\Lambda $ is the frequency and the lattice field $u_{\mathbf{n}}$
obeys the equation
\begin{equation}
(\Lambda -\sigma |u_{\mathbf{n}}|^{2})u_{\mathbf{n}}-\varepsilon \Delta u_{%
\mathbf{n}}=0.  \label{schr2}
\end{equation}

Our presentation is structured as follows. The following section
recapitulates the necessary background for the prediction of the existence
and stability of lattice solitons. In section III, we report a bifurcation
analysis for various surface states, treated as functions of coupling
constant $\varepsilon$, with emphasis on the comparison with bulk
counterparts of these states. Section IV reports the study of the evolution
of unstable surface states. Finally, section V summarizes our findings and
presents our conclusions.

\section{The theoretical background}

\label{sec:TheorBack}

First, we outline some general properties of the model. Equation (\ref{schr1}%
) conserves two dynamical invariants, namely the norm $N$,
\begin{equation}
N=\sum_{\substack{ {n_{3}=1}  \\ {n_{1,2}=-\infty }}}^{\infty }|\phi _{%
\mathbf{n}}|^{2},
\end{equation}
and the Hamiltonian $H$,
\begin{equation}
H=\sum_{\substack{ {n_{3}=1}  \\ {n_{1,2}=-\infty }}}^{\infty }\left(
\varepsilon \sum_{j=1}^{3}\left[ \phi _{\mathbf{n}}^{\ast }(\phi _{\mathbf{n}%
+\mathbf{e}_{j}}-s\phi _{\mathbf{n}})+\mathrm{c.c.}\right] +\frac{\sigma }{2}%
|\phi _{\mathbf{n}}|^{4}\right) ,
\end{equation}
where the asterisk stands for complex conjugation. Stationary solutions to
Eq.~(\ref{schr2}) with $\sigma =\pm 1$ are connected by the \textit{%
staggering transformation}~\cite{ABK,BluKon}: if $u_{\mathbf{n}}$ is a
solution for some $\Lambda $ and $\sigma =+1$, then $%
(-1)^{n_{1}+n_{2}+n_{3}}u_{\mathbf{n}}$ is a solution for $\widetilde{%
\Lambda }=12s-\Lambda $ and $\sigma =-1$. Consequently, it is sufficient to
perform the analysis of stationary solutions, including their stability, for
a single sign of the nonlinearity; thus, below we will fix $\sigma =+1$
(corresponding to the case of onsite self-attraction).

Solutions to Eq.~(\ref{schr2}) in half-space $n_{3}\geq 1$, subject to
boundary condition $\phi _{\mathbf{n}}=0$ for $n_{3}=0$, as defined above,
may be continued anti-symmetrically for the entire 3D space by setting $U_{%
\mathbf{n}}\equiv u_{\mathbf{n}}$ for $n_{3}\geq 1$ and $U_{\mathbf{n}%
}\equiv -u_{\mathbf{n}}$ for $n_{3}\leq -1$. Then, according to results of
Ref.~\cite{Weinstein}, this leads to an immediate conclusion, namely that
there exists a minimum norm $N_{\min }$ necessary for the existence of
localized surface states in the present model. In other words, no surface
modes survive in the limit of $N\rightarrow 0$. In this connection, it is
relevant to note that numerical findings that will be presented below were
obtained, of course, for finite cubic lattices where, strictly speaking,
there is no lower limit for $N$ necessary for the existence of localized
modes~\cite{BluKon}. At this point, we have to specify that speaking about
localized modes in a finite lattice we understand solutions which are
localized on a number of cites much smaller than the total number of sites
in the chosen direction used for numerical simulations. Next we recall that
generally speaking, there exist several branches of the nonlinear localized
modes, i.e. for a given $\varepsilon$ one can find localized excitations at
different values of the norm $N$. Using the natural terminology we refer to
higher/lower branches speaking about solutions with larger/smaller norm. In
this classification the surface modes we are dealing with correspond to
higher branches of the solutions of the respective finite lattices, i.e.,
their norm cannot be made arbitrarily small (see also the relevant
discussion below in Section \ref{small-amp}).

To find solution families, we start with the anti-continuum (AC) limit, $%
\varepsilon =0$ \cite{Pelinovsky}. In this limit, the lattice field is
assumed to take nonzero values only at a few (``excited") sites, which
determines the profile of the configuration to be seeded. The continuation
of the structure to $\varepsilon >0$ is determined by the Lyapunov's
reduction theorem \cite{Golubitsky}. More specifically, the solution is
expanded as a power series in $\varepsilon $, the solvability condition at
each order being that the respective projection to the kernel generated by
the previous order does not give rise to secular terms \cite{Pelinovsky}.

The linear stability is then studied, starting from the usual form of the
perturbed solution,
\begin{equation}
\phi _{\mathbf{n}}=e^{i\Lambda t}(u_{\mathbf{n}}+\delta a_{\mathbf{n}%
}e^{\lambda t}+\delta b_{\mathbf{n}}e^{\lambda ^{\ast }t}),
\end{equation}
where $\delta $ is a formal small parameter, and $\lambda $ is a stability
eigenvalue associated with eigenvector $\psi =\{a_{\mathbf{n}},b_{\mathbf{n}%
}^{\ast }\}$. Substituting this into Eq.~(\ref{schr1}) yields the linearized
system,
\begin{eqnarray}
i\lambda a_{\mathbf{n}} &=&-\varepsilon \Delta a_{\mathbf{n}}+\Lambda a_{%
\mathbf{n}}-2|u_{\mathbf{n}}|^{2}a_{\mathbf{n}}-u_{\mathbf{n}}^{2}b_{\mathbf{%
n}}^{\ast }\,,  \notag \\[2ex]
-i\lambda b_{\mathbf{n}}^{\ast } &=&-\varepsilon \Delta b_{\mathbf{n}}^{\ast
}+\Lambda b_{\mathbf{n}}^{\ast }-2|u_{\mathbf{n}}|^{2}b_{\mathbf{n}}^{\ast }-%
{u_{\mathbf{n}}^{\ast }}^{2}a_{\mathbf{n}}\,,  \notag
\end{eqnarray}
which can be cast in the form
\begin{equation}
\left(
\begin{array}{ccc}
~\mathcal{H}^{(1,1)} &  & ~\mathcal{H}^{(1,2)} \\[1ex]
\mathcal{H}^{(2,1)} &  & ~\mathcal{H}^{(2,2)}%
\end{array}%
\right) \left(
\begin{array}{c}
\mathcal{A} \\[1ex]
\mathcal{B}%
\end{array}%
\right) =i\lambda \left(
\begin{array}{c}
\mathcal{A} \\[1ex]
\mathcal{B}%
\end{array}%
\right) ,  \label{eig-prob}
\end{equation}
where $\mathcal{A}$ and $\mathcal{B}$ are vectors composed by elements $a_{%
\mathbf{n}}$ and $b_{\mathbf{n}}^{\ast }$, respectively, while the matrices $%
\mathcal{H}^{(p,q)}$ ($p, q \in\{1,2\}$) are given by,
\begin{eqnarray}
\mathcal{H}_{\mathbf{n},\mathbf{n}^{\prime }}^{(1,1)} &=&-\mathcal{H}_{%
\mathbf{n},\mathbf{n}^{\prime }}^{(2,2)}=\delta _{\mathbf{n},\mathbf{n}%
^{\prime }}\left( \Lambda +6s\varepsilon -2|u_{\mathbf{n}^{\prime
}}|^{2}\right)  \notag \\
&&-\varepsilon \sum_{j=1}^{3}\left( \delta _{\mathbf{n}+\mathbf{e}_{j},%
\mathbf{n}^{\prime }}+\delta _{\mathbf{n}-\mathbf{e}_{j},\mathbf{n}^{\prime
}}\right) ,  \label{energy2d} \\
\mathcal{H}_{\mathbf{n},\mathbf{n}^{\prime }}^{(1,2)} &=&-{\mathcal{H}_{%
\mathbf{n},\mathbf{n}^{\prime }}^{(2,1)}}^{\ast }=-\delta _{\mathbf{n},%
\mathbf{n}^{\prime }}u_{\mathbf{n}^{\prime }}^{2}.  \notag
\end{eqnarray}%
An underlying stationary solution is (spectrally) unstable if there exists a
solution to Eq.~(\ref{eig-prob}) with $\mathrm{Re}(\lambda )>0$. Otherwise,
the stationary solution is classified as a spectrally stable one. As
explained in Ref.~\cite{Pelinovsky1d}, the Jacobian of the above mentioned
solvability conditions is intimately connected to the full eigenvalue
problem. More specifically, if the eigenvalues $\gamma$ of the $M \times M$
eigenvalue problem of the Jacobian (where $M$ is the number of excited sites
at the AC limit), then the near-zero eigenvalues of the full stability
problem can be predicted to be $\lambda=\sqrt{2 \gamma} \varepsilon^{p/2}$,
where $p$ is the number of lattice sites that separate the adjacent excited
nodes of the configuration at the AC limit.


\section{The bifurcation analysis}


\subsection{Existence and stability of surface structures}

In this section we study, by means of numerical methods, the existence and
stability of various 3D configurations and compare the results to the
corresponding analytical predictions. These configurations are obtained by
starting from the AC limit $(\varepsilon =0)$, and are continued to $%
\varepsilon >0$, using fixed-point iterations. For all the numerical results
presented in this work, we fix the normalization $\Lambda =1$ [see Eq.~(\ref%
{schr2})], and use a lattice of size $13\!\times 13\!\times 13$, unless
stated otherwise. Also, for the presentation of the numerical results, we
replace the triplet $(n_1,n_2,n_3)$ by $(l,n,m)$, i.e., the surface
corresponds to $m=1$.

\begin{figure}[thbp]
\includegraphics[width=\figw,height=3.9cm,angle=0]{\rootfig 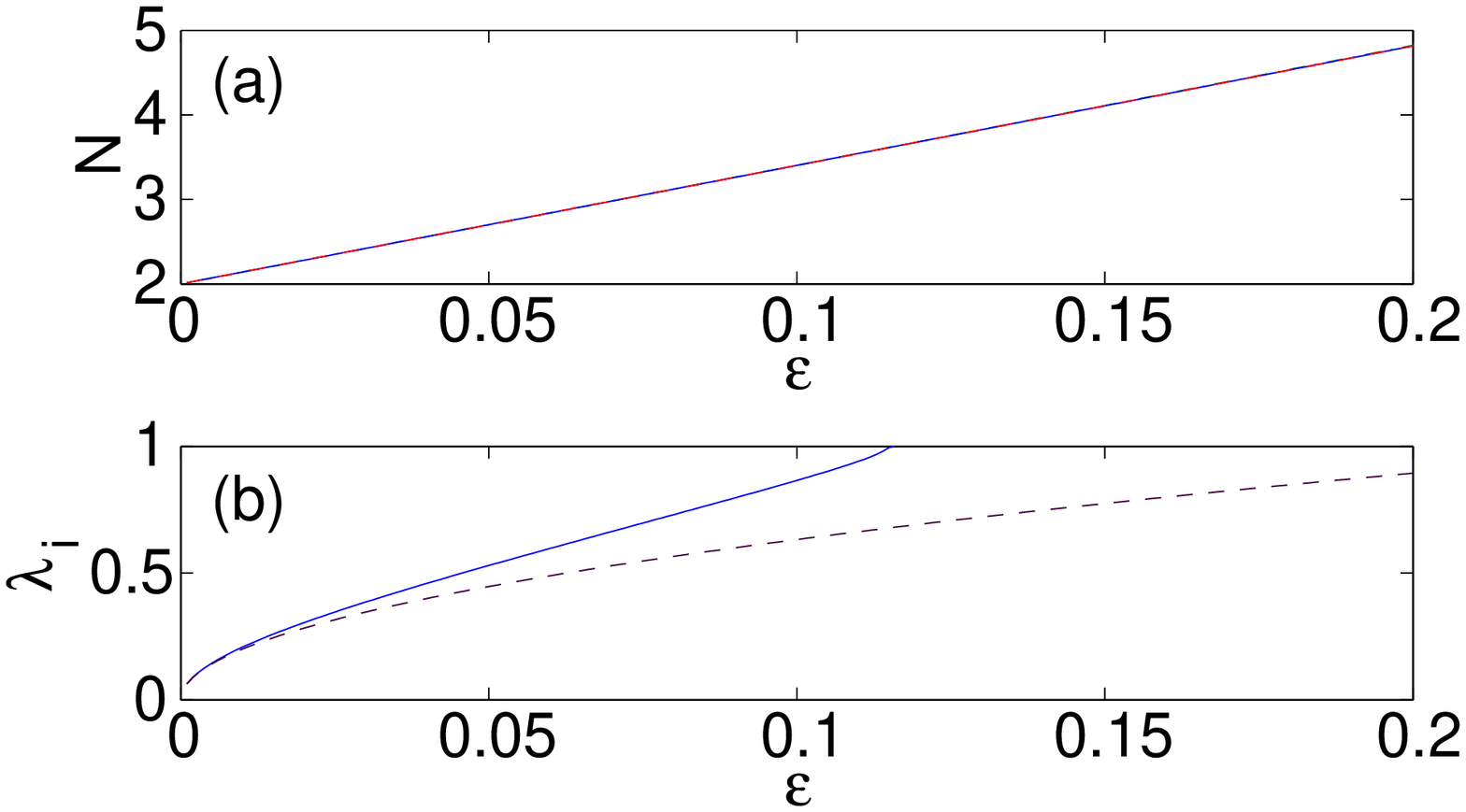}\\[%
-1.5ex]
\includegraphics[width=\figw,height=3.9cm,angle=0]{\rootfig 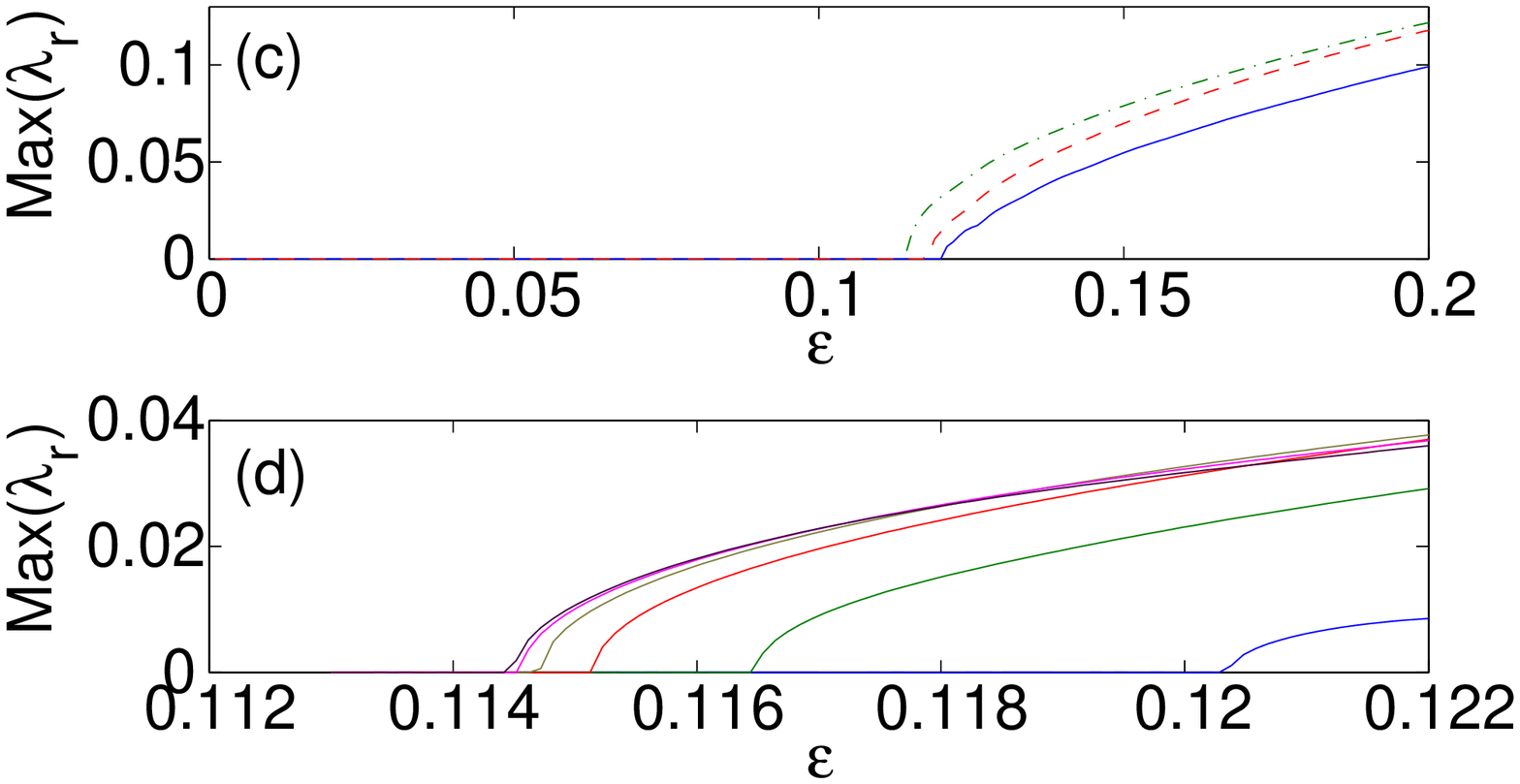}%
\\[-1.5ex]
\includegraphics[width=\figw,height=2.2cm,angle=0]{\rootfig 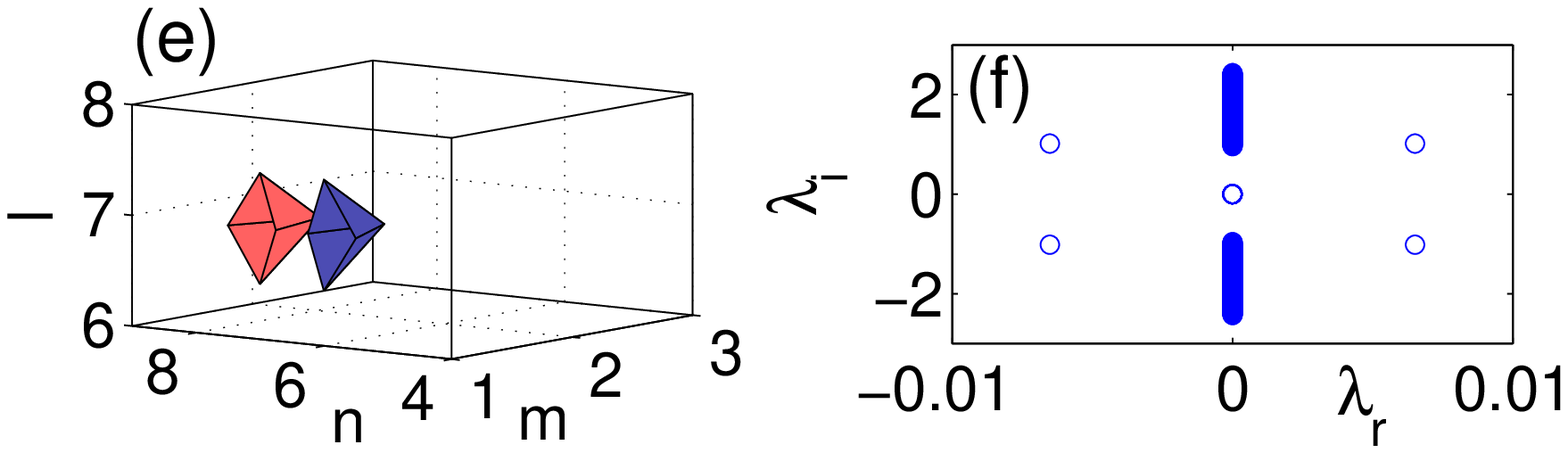}\\[%
0.0ex]
\includegraphics[width=\figw,height=2.2cm,angle=0]{\rootfig 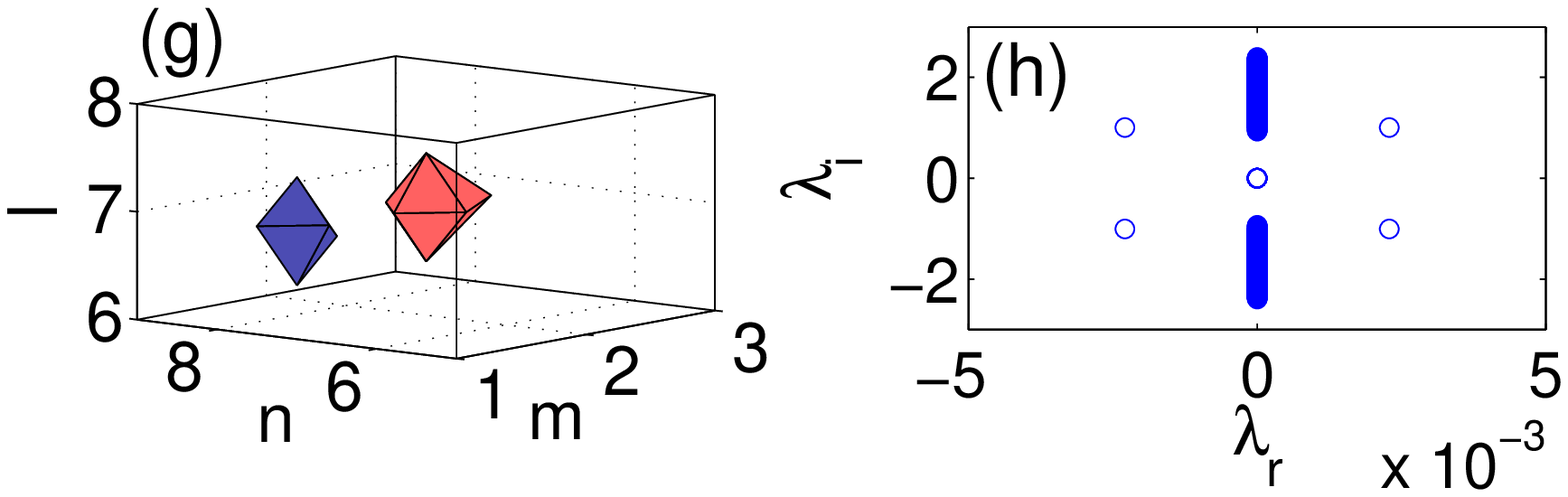}
\caption{(Color Online) Results for the dipoles oriented parallel and normal
to the surface. (a) Norm $N$ versus the lattice coupling
constant, $\protect\varepsilon $. (b) Imaginary part
of the linear stability eigenvalue: solid (blue) and dashed (black) lines
correspond, respectively, to numerically found and analytically predicted
forms. (c) Real part of the critical (in)stability
eigenvalue: the dashed (red) and solid (blue) lines depict the normal- and
parallel-oriented dipoles, respectively, while the dash-dotted (green) line
corresponds to the bulk dipole. (d) (In)stability
eigenvalue for the parallel surface dipole placed at distances from the
surface starting from zero and up to five lattice periods away (curves right
to left). (e)-(g) Configurations and (f)-(h) respective
spectral stability planes just above the
instability threshold. The level contours, corresponding to $\mathrm{Re}%
(u_{l,n,m})=\pm 0.5\max \left\{ u_{l,n,m}\right\} $ are shown, respectively,
in dark gray (blue) and gray (red). The instability thresholds for the
dipoles oriented parallel and normally to the surface are, respectively, $%
\protect\varepsilon =0.117$ and $\protect\varepsilon =0.120$. For
comparison, the threshold for the bulk dipole is $\protect\varepsilon =0.114$%
. }
\label{Fig_dip}
\end{figure}

We start by examining dipoles aligned parallel or normal to the surface.
Panel (a) in Fig.~\ref{Fig_dip} shows the norm of such states versus
coupling constant $\varepsilon $, while panel (b) depicts the
imaginary part of the stability eigenvalues for the bulk dipole, produced by
the theory outlined in the previous section [dashed (black) lines], and by
the numerical computations [solid (blue) lines]. It is worth mentioning
that, for all the different configurations that we report in this
manuscript, we display the imaginary part of the stability eigenvalue only
for the \emph{bulk} mode since the difference between the curves for the
different variants (bulk, parallel or normal to the surface) is minimal. It
should be noted however that the contact with the surface may produce higher
order (smaller) eigenvalues that are not present in its bulk counterpart
(results not shown here). The theoretical prediction for the stability
eigenvalues is $\lambda =\pm 2\sqrt{\varepsilon }i$, which, as expected, is
the same as in an out-of-phase (\textit{twisted}) 1D mode analyzed in Ref.~%
\cite{Pelinovsky1d}, since the structure is nearly one-dimensional, along
the line connecting the two excited sites.
%
Panel (c) in Fig.~\ref{Fig_dip} compares the largest instability
growth rate as a function of $\varepsilon $ for the bulk dipoles
[dash-dotted (green) line] and those oriented normally and parallel to the
surface [dashed (red) and solid (blue) lines, respectively]. It is seen that
the stability interval of the dipoles increases as its contact with the
surface strengthens, in accordance with the arguments presented above. In
the case of the bulk dipole, the instability occurs for values of the
coupling constant in between $\varepsilon_{0}=0.114$ and $\varepsilon
_{1}=0.115$. From now on, when reporting computed instability thresholds, we
will use the lower bound for $\varepsilon $ (e.g., $\varepsilon_{0}$ in the
above example) with the understanding that we always used the same
resolution in $\varepsilon$.
For the dipole set normally to the surface, we observe the onset of
instability at $\varepsilon =0.117$, while for the parallel-oriented one at $%
\varepsilon =0.120$. In panels (e)--(h) of Fig.~\ref{Fig_dip} we also depict 
the shapes of the normal and parallel dipoles, just
below the instability threshold, along with their corresponding spectral
stability planes.

The stabilizing effects exerted by the surface depend, in a great measure,
on the distance of the configuration from the surface, namely, the further
away the configuration from the surface, the lesser the effect is. This
property is clearly seen in panel (d) of Fig.~\ref{Fig_dip}, where we
plot the (in)stability eigenvalue as a function of the coupling for several
values of the separation of the parallel dipole from the surface. The
curves, from right to left, depict the results for the dipole set at the
distance of 0, 1, ..., 5 sites away from the surface (0 sites refers to the
dipole sitting on the surface). As the panel demonstrates, the stability
interval is reduced as the dipole is pulled away from the surface,
converging towards a bulk dipole.

\begin{figure}[tp]
\centering
\includegraphics[width=\figw,height=3.9cm,angle=0]{\rootfig 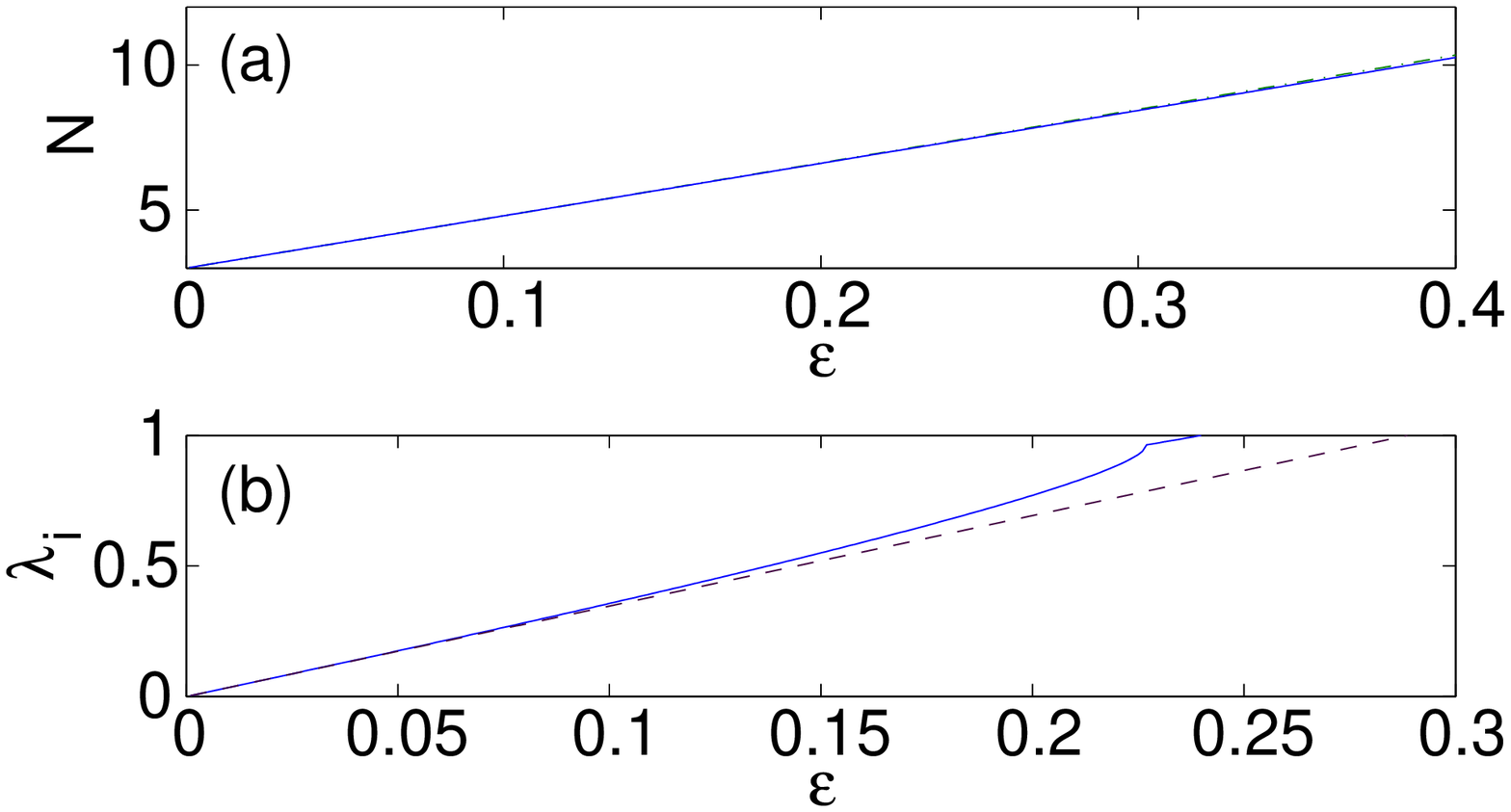}
\\[-2.0ex]
\includegraphics[width=\figw,height=2.1cm,angle=0]{\rootfig 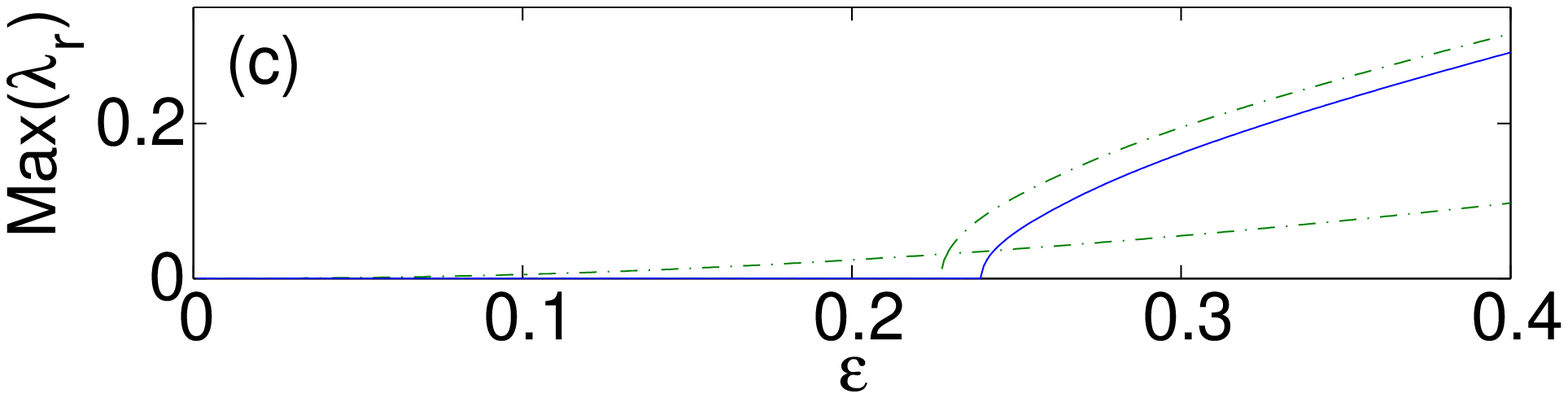}
\\[-1.0ex]
\includegraphics[width=\figw,height=2.2cm,angle=0]{\rootfig 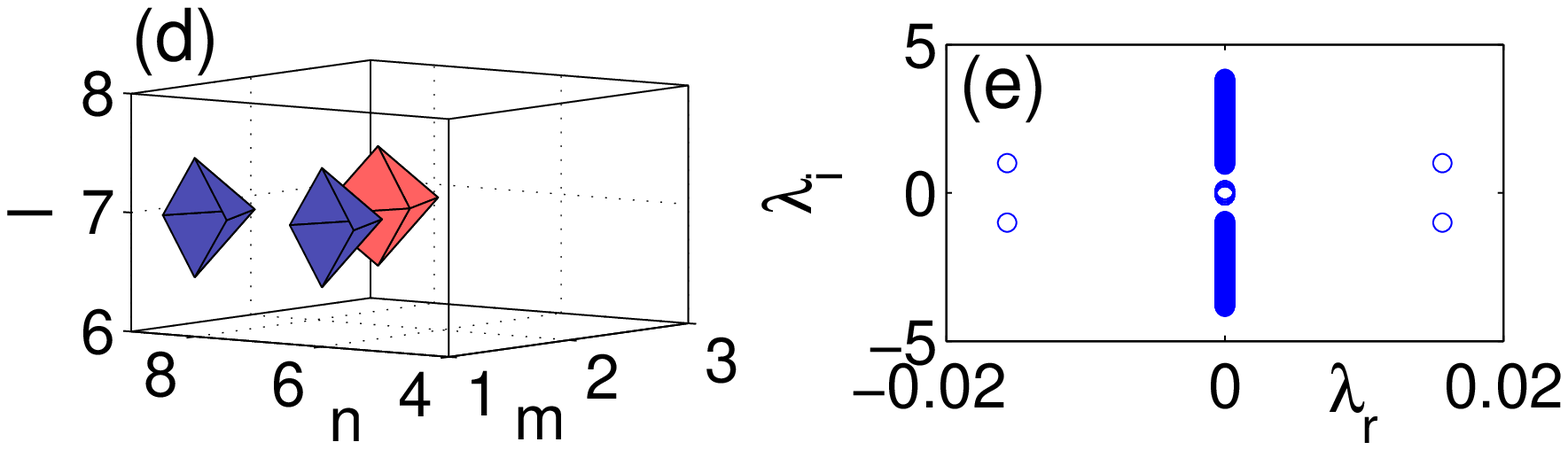}
\caption{(Color Online) The stability of the three-site ``horseshoe''.
Panels are similar to those in Fig.~\protect\ref{Fig_dip}. Panel (c)
compares the critical stability eigenvalue, as a function of the lattice
coupling, $\protect\varepsilon $, for the surface and bulk horseshoes [solid
(blue) and dashed-dotted (green) lines, respectively]. The bulk horseshoe is
always unstable (due to a purely real, higher-order eigenvalue), while the
corresponding surface configurations have a stability region (the
corresponding eigenvalue becomes imaginary in this case). Panels
(d)-(e) correspond to the surface horseshoe just above the stability
threshold of $\protect\varepsilon =0.239$.}
\label{Fig_3_horse}
\end{figure}

Let us now consider the ``horseshoe" configurations, for which the presence
of the surface is critical to their stability. In Fig.~\ref{Fig_3_horse} we
depict the properties of a three-site horseshoe, which actually is a
truncated version of a quadrupole, cf. the 2D situation \cite{Susanto}. As
before, panel (a) in Fig.~\ref{Fig_3_horse} shows the norm versus $%
\varepsilon $, while panels (b) and (c) compare the stability of the
bulk horseshoe (the dash-dotted line) and ones built near the surface (the
solid line). While the bulk horseshoes are always unstable, similar to their
2D counterparts \cite{Susanto}, the ones placed near the surface are stable
at small $\varepsilon$, destabilizing at $\varepsilon =0.239$. Panels (d)-(e)
in Fig.~\ref{Fig_3_horse} show the configuration for the
coupling just below the instability threshold, along with the respective
spectral plane. The analytical expressions for stable eigenvalues are $%
\lambda =0$, $\lambda =\pm 2\sqrt{3}\varepsilon i$, $\lambda =\mathcal{O}%
(\varepsilon ^{2})$, \ cf.~the expressions obtained in Ref.~\cite{Susanto}
for the 2D horseshoes.

\begin{figure}[tbp]
\centering
\includegraphics[width=\figw,height=3.9cm,angle=0]{\rootfig 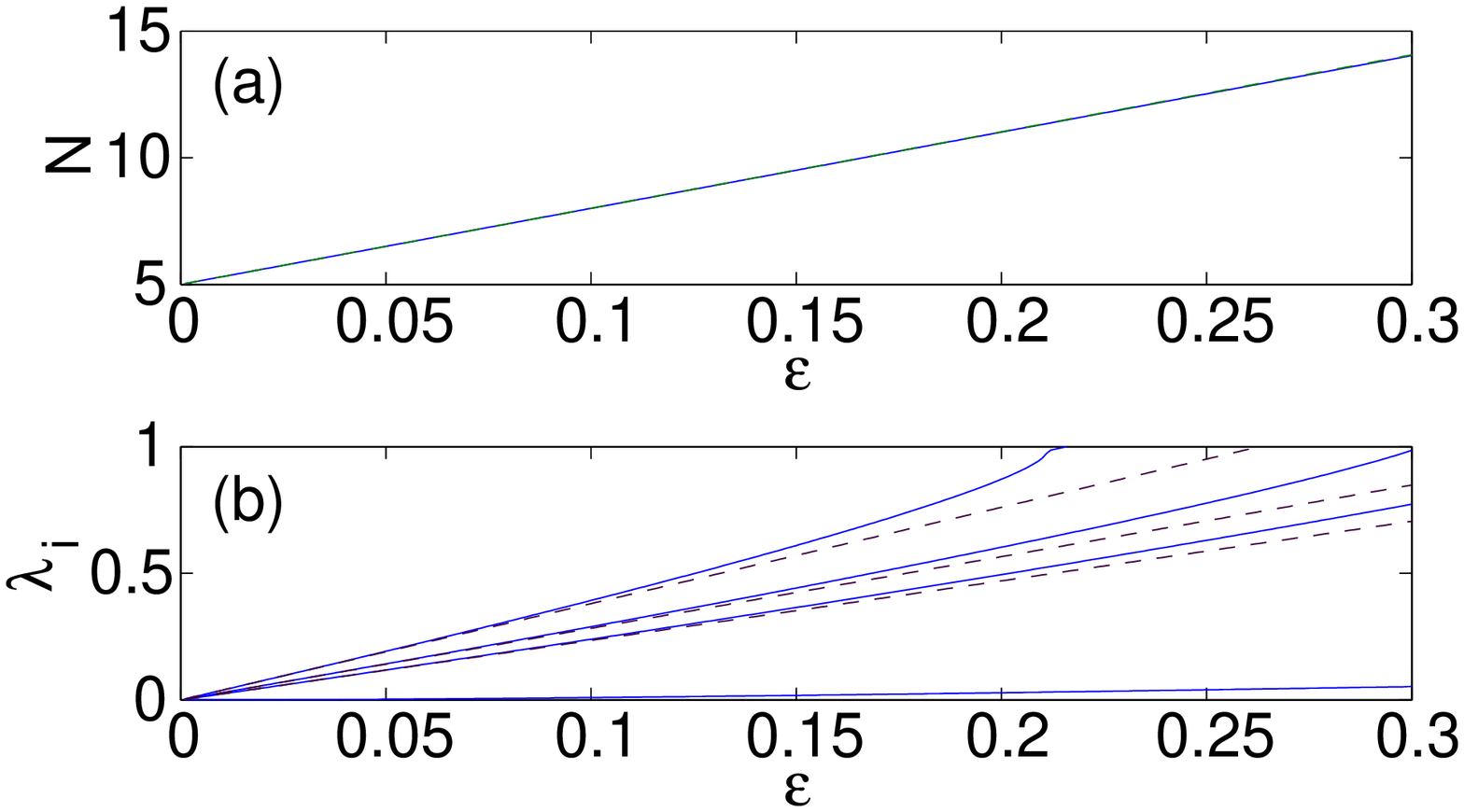}
\\[-2.0ex]
\includegraphics[width=\figw,height=2.0cm,angle=0]{\rootfig 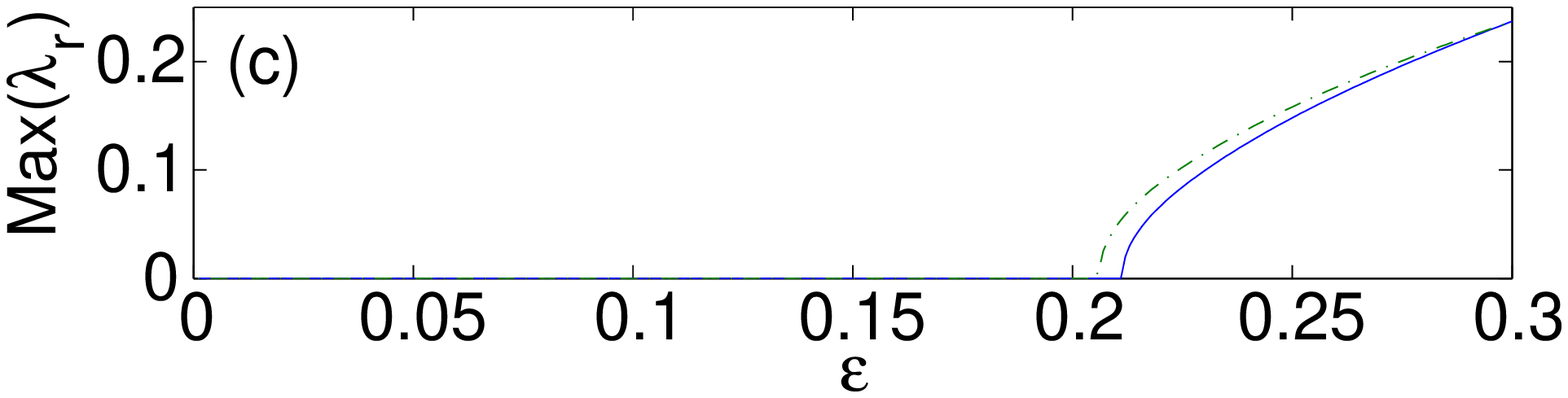}
\\[-1.0ex]
\includegraphics[width=\figw,height=2.2cm,angle=0]{\rootfig 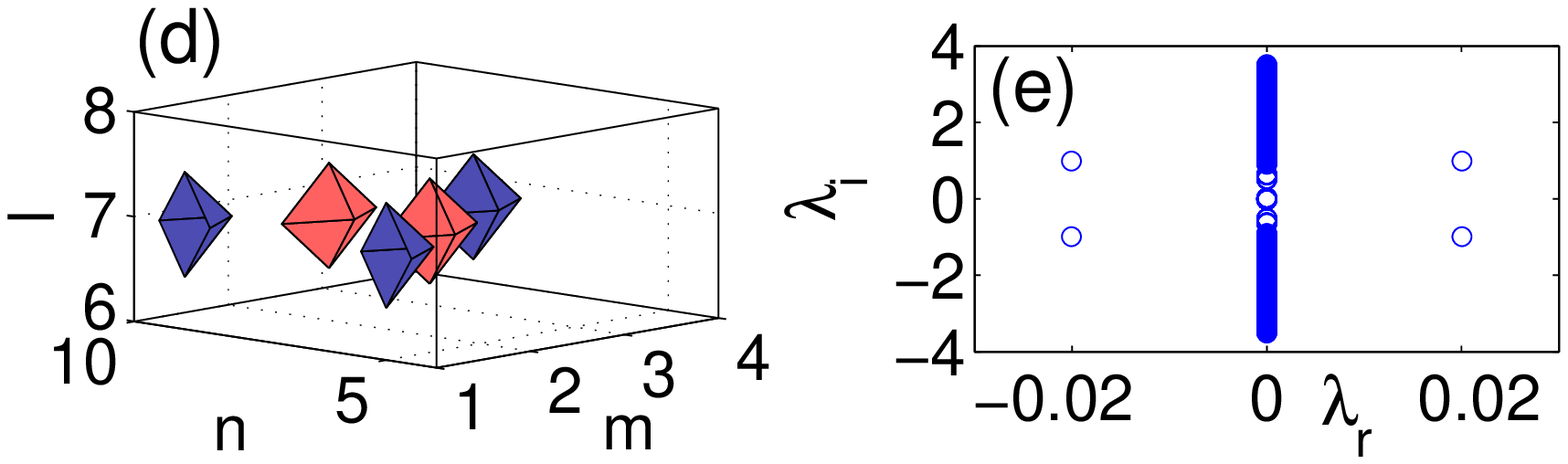}
\caption{(Color Online) The stability for the five-site horseshoe at the
surface. Panels are identical to those in Fig.~\protect\ref{Fig_3_horse}. In
this case, the stability threshold is at $\protect\varepsilon =0.211$, while
for the bulk 5-site horseshoe it is $\protect\varepsilon =0.205$.
Panels (d)-(e) depict the configuration and the respective linear stability
spectrum just above the critical point of $\protect\varepsilon =0.211$.}
\label{Fig_5_horse}
\end{figure}

Figure~\ref{Fig_5_horse} illustrates the same features as before but for the
five-site horseshoe. Unlike its three-site cousin, the bulk five-site
horseshoe is \emph{stable} up to a critical value of the coupling, $%
\varepsilon =0.205$, while the surface variant has it stability region $%
\varepsilon <0.211$. The eigenvalues of the linearization in this case can
be computed similar to those for the three-site modes \cite{Susanto}, as
outlined above (cf.~also Ref.~\cite{Pelinovsky}), which eventually yields $%
\lambda =3.8042\varepsilon i$, $\lambda =2.8284\varepsilon i$, $\lambda
=2.3511\varepsilon i$, $\lambda =\mathcal{O}(\varepsilon ^{2})$, and $%
\lambda =0$, in good agreement with the corresponding numerical results, as
shown in panel (b) Fig.~\ref{Fig_5_horse}.

\begin{figure}[tbp]
\centering
\includegraphics[width=\figw,height=3.9cm,angle=0]{\rootfig 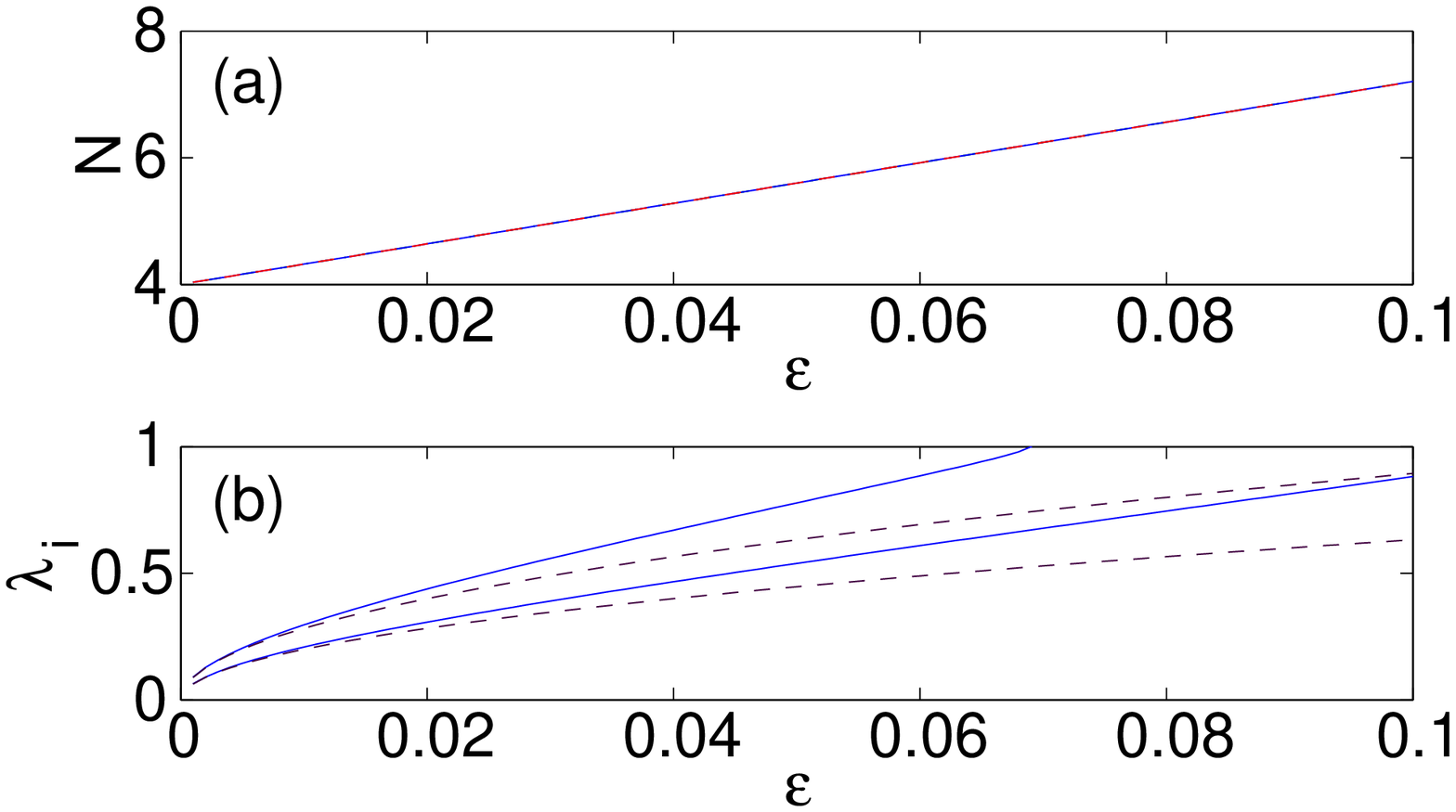}
\\[-2.0ex]
\includegraphics[width=\figw,height=2.0cm,angle=0]{\rootfig 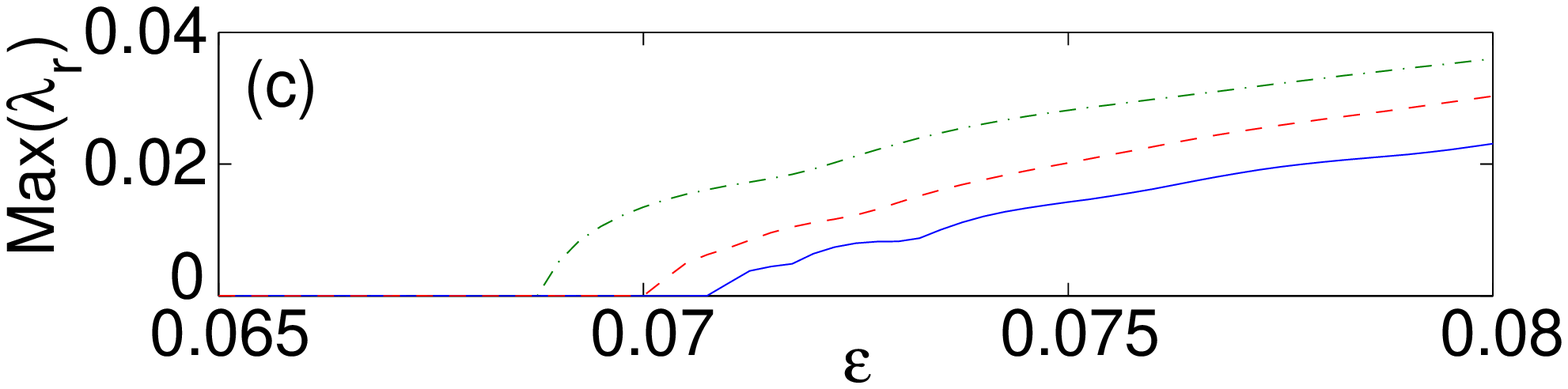}
\\[-0.0ex]
\includegraphics[width=\figw,height=2.2cm,angle=0]{\rootfig 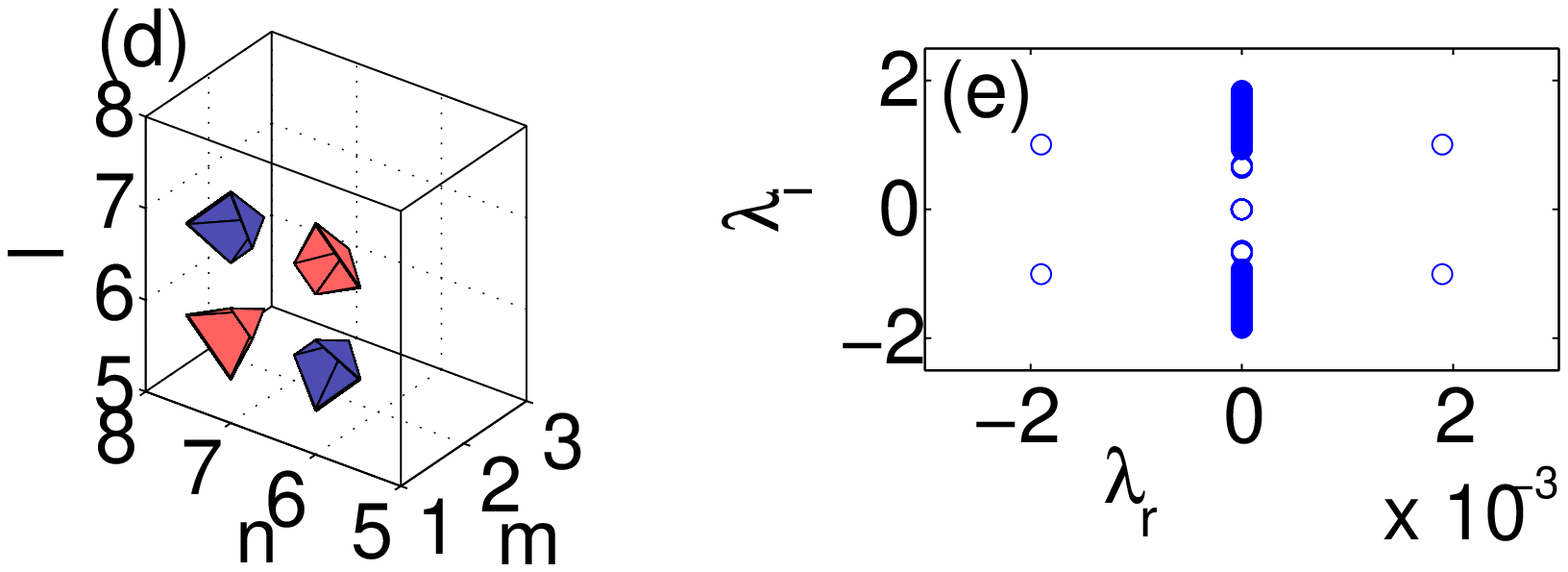}
\caption{(Color Online) The stability of quadrupole modes. The layout is
similar to that in Fig.~\protect\ref{Fig_5_horse}. In panel (c), due
to the close proximity of the thresholds, the close-up is shown for the
critical stability eigenvalue versus the lattice coupling constant, $\protect%
\varepsilon $, for the parallel and normal surface modes, and the bulk one
[solid (blue) and dashed (red) lines, and the dash-dotted (green) line,
respectively]. The threshold for the bulk mode is $\protect\varepsilon %
=0.068 $, while for the normal and parallel quadrupoles it is, respectively,
$\protect\varepsilon =0.070$ and $\protect\varepsilon =0.071$. As before,
panels (d) and (e) show the configuration just above the instability
threshold along with its corresponding spectral-stability plane. }
\label{Fig_quad}
\end{figure}

Next we consider the quadrupole configuration, see Fig.~\ref{Fig_quad}. The
surface again exerts a stabilizing effect, albeit a weaker one, when the
quadrupole is placed normally and parallel to the surface. In the bulk, the
quadrupole loses stability at $\varepsilon =0.068$, while the normal and
parallel surface quadrupoles have stability thresholds, respectively, at $%
\varepsilon =0.070$ and $\varepsilon =0.071$. The analytical approximation
for the stability eigenvalues in this case are $\lambda =\sqrt{8\varepsilon }%
i$ (a double eigenvalue), $\lambda =2\sqrt{\varepsilon }i$, and a zero
eigenvalue, which accurately capture the numerical findings depicted in panel (b)
of Fig.~\ref{Fig_quad}.

\begin{figure}[tbp]
\centering
\includegraphics[width=\figw,height=1.9cm,angle=0]{\rootfig 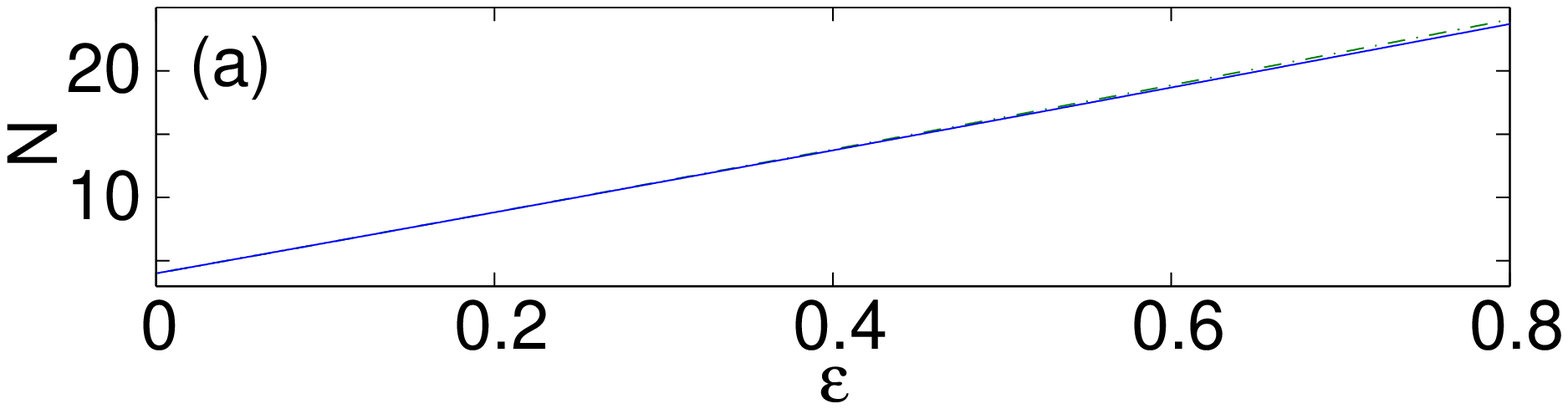} \\[%
-1.0ex]
\includegraphics[width=\figw,height=1.9cm,angle=0]{\rootfig 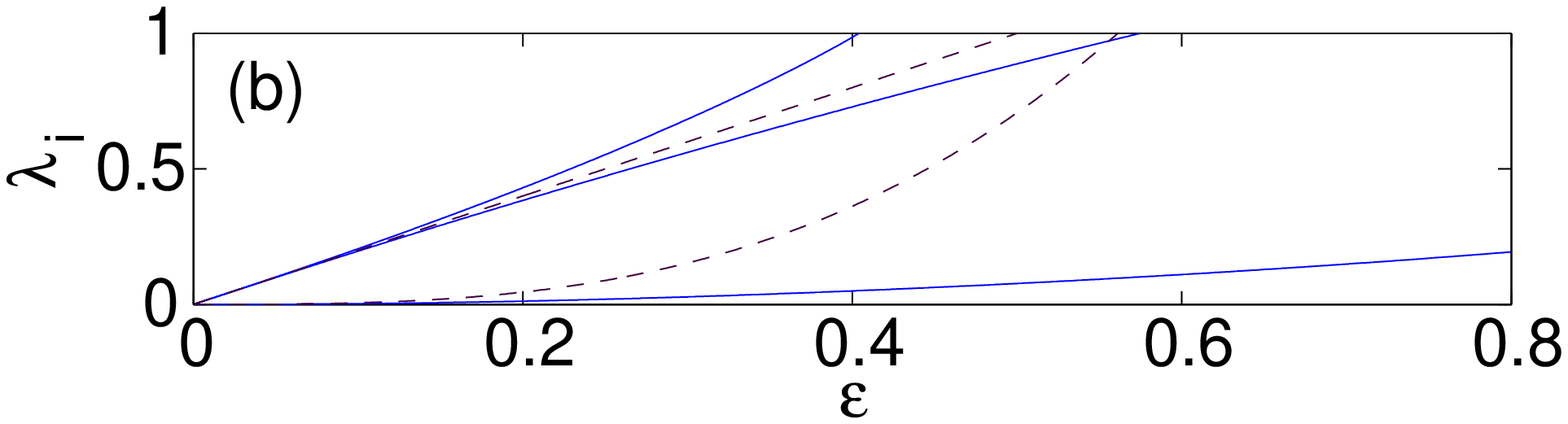} \\[%
-1.0ex]
\includegraphics[width=\figw,height=1.9cm,angle=0]{\rootfig 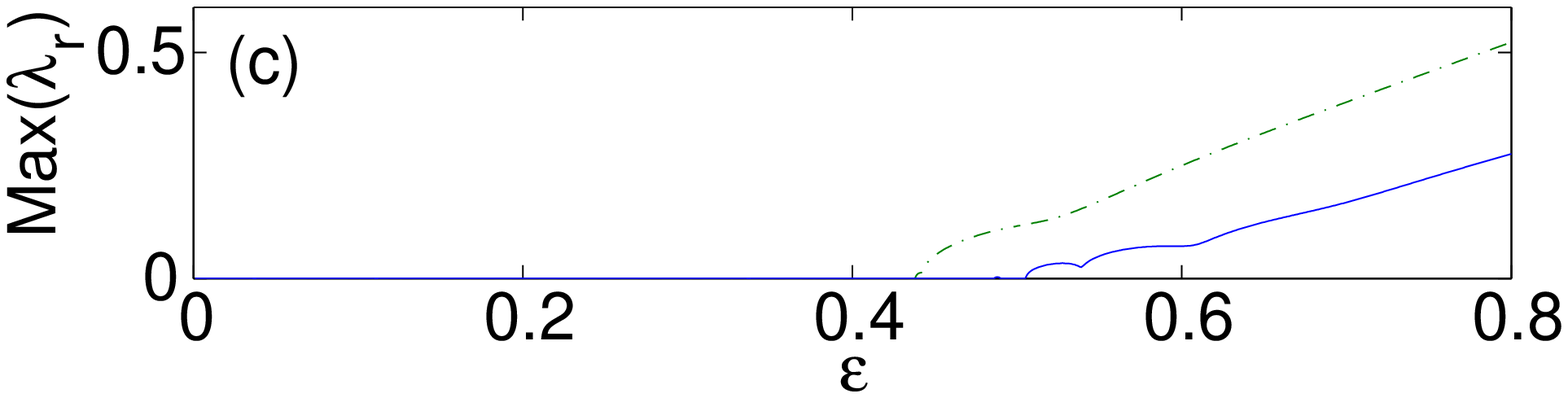} \\[%
-1.0ex]
\includegraphics[width=\figw,height=2.1cm,angle=0]{\rootfig 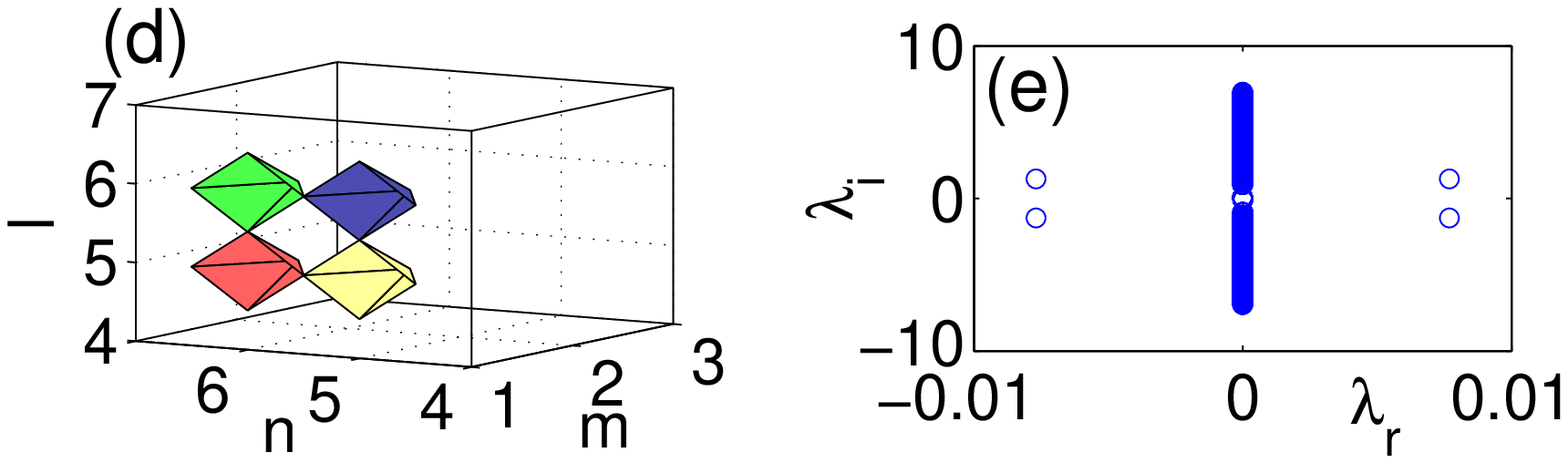}
\caption{(Color Online) The stability of the four-site vortex in the grid of
size $11\!\times 11\!\times 11$. The dash-dotted and solid lines show the
bulk vortex and the one parallel to the surface, respectively. The layout is
similar to that of the above figures. Instability in the bulk occurs at $%
\protect\varepsilon =0.438$, and in the parallel surface vortex at $\protect%
\varepsilon =0.505$. The vortex cannot exist with the orientation normal to
the surface. Panels (d) and (e) show the parallel surface vortex just
above the instability threshold of $\protect\varepsilon =0.485$. As in the
previous figures, the level contours corresponding to $\mathrm{Re}%
(u_{l,n,m})=\pm 0.5\max \left\{ u_{l,n,m}\right\} $ are shown, respectively,
in dark gray (blue) and gray (red), while the complementary level contours,
defined as $\mathrm{Im}(u_{l,n,m})=\pm 0.5\max \left\{ u_{l,n,m}\right\} $,
are shown by light gray (green) and very light gray (yellow) hues,
respectively.}
\label{Fig_vort}
\end{figure}

In Fig.~\ref{Fig_vort} we present the results for four-site vortices. This
configuration, in contrast to the previous ones, is described by a complex
solution. In the AC limit, the vortex occupies the same excited sites as the
above-mentioned quadrupole, but the phase profile, $\{0,\pi /2,\pi ,3\pi
/2\} $, emulates that of the vortex of charge $1$ \cite{Malomed,Pelinovsky}.
The bulk four-site vortex (which was discussed in Ref.~\cite{earlier3})
loses its stability at $\varepsilon =0.438$, while the vortex parallel to
the surface features an extended stability region, up to $\varepsilon =0.505$.
However, the surface in this case \emph{prohibits} the existence of a
vortex that would be oriented normally to the surface layer, similarly to
what was found for 2D lattice vortices~\cite{Susanto}.


The simplest explanation for the complete absence of the solution 
normal to the surface, compared with that of an existing vortex
waveform parallel to the surface can arguably be traced in the interaction
of such vortices in the half-space with their fictitious image
(if the domain is equivalently extended to the full space). In the case of
the vortex parallel to the surface, the situation is tantamount
to the vortex cube structures examined in
\cite{vor3dnew,Pelinovsky3d},
for which it was established in \cite{Pelinovsky3d} that the
persistence conditions are satisfied (and, in fact, that such
structures consisting of two out-of-phase vortices should
be linearly stable close to the AC limit). On the other hand,
for the case normal to the surface, by examining the relevant
interactions it can be observed (at an appropriately high order) that
the persistence conditions of
\cite{Pelinovsky,Pelinovsky1d,Pelinovsky3d} can not
be satisfied and hence the structure can not be continued beyond the
AC limit. That is why the structure can never be observed to exist
irrespectively of the smallness of $\varepsilon$.

\begin{figure}[tbp]
\centering
\includegraphics[width=\figw,height=2cm,angle=0]{\rootfig 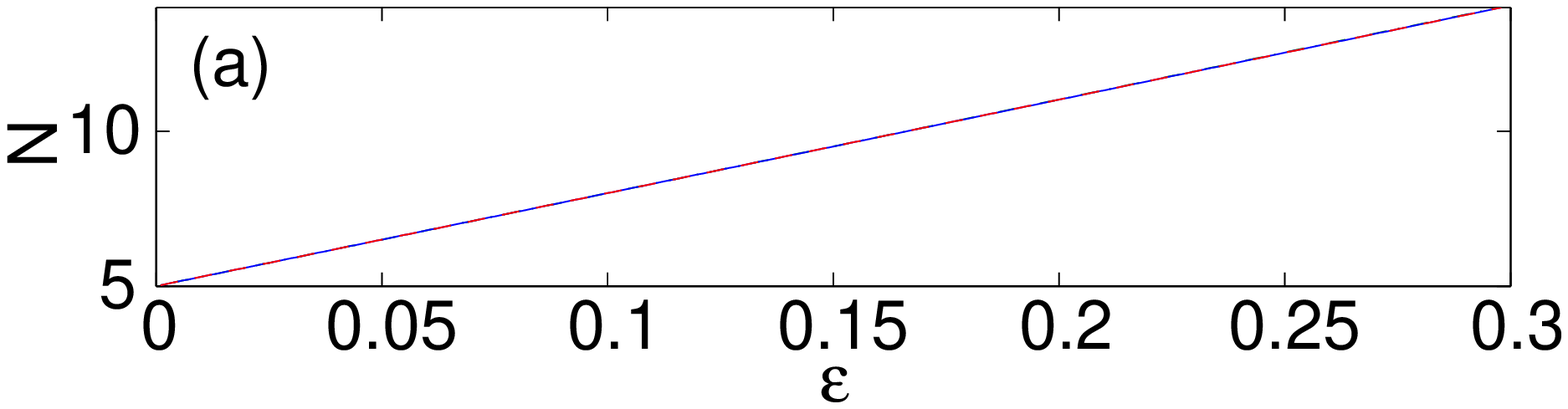} \\[%
-1.0ex]
\includegraphics[width=\figw,height=2cm,angle=0]{\rootfig 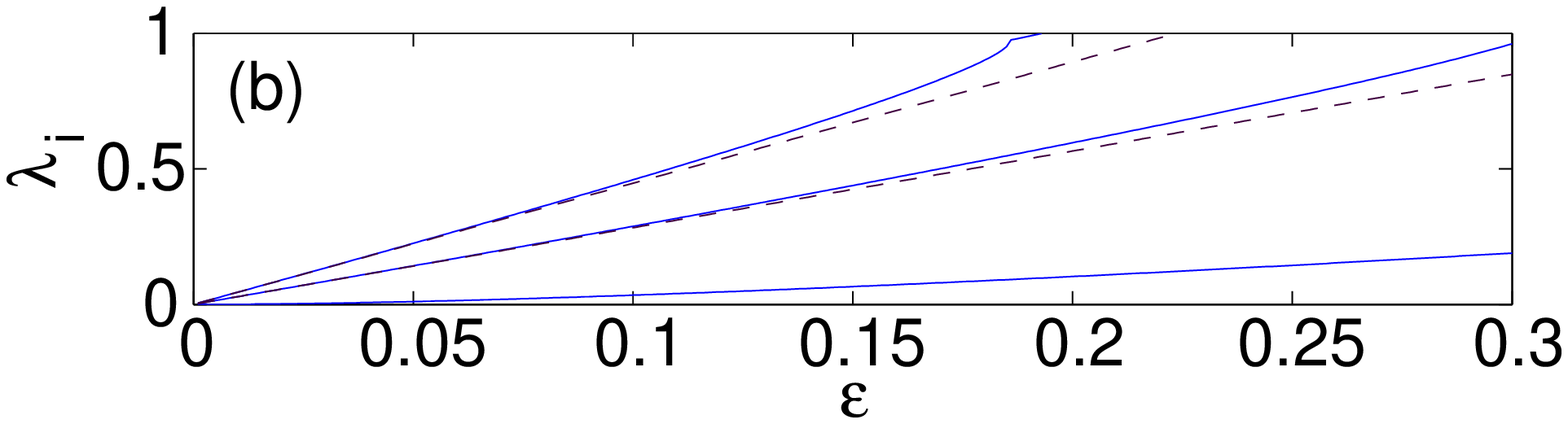} \\[%
-1.0ex]
\includegraphics[width=\figw,height=2cm,angle=0]{\rootfig 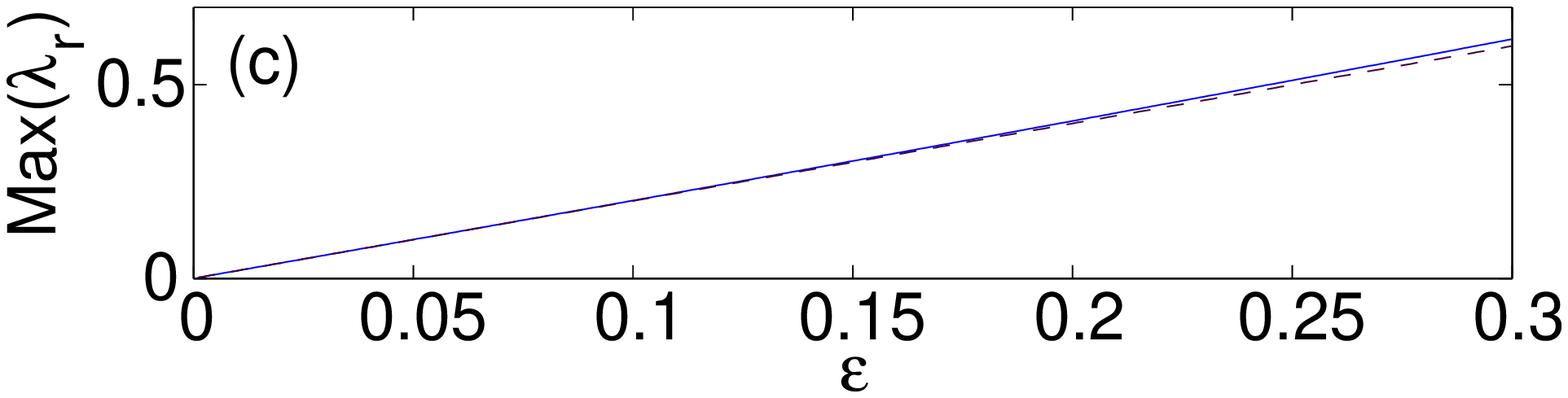} \\[%
1.0ex]
\includegraphics[width=8.2cm,height=2.2cm,angle=0]{\rootfig 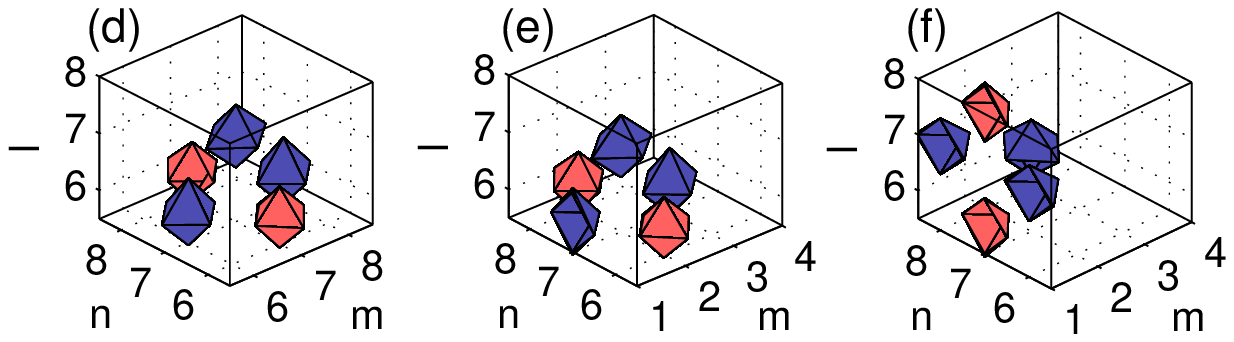}
\caption{(Color Online) The instability of pyramid-shaped structures. This
configuration abuts on the base in the form of a rhombus, and includes the
out-of-plane site with zero phase. Three variants of this configuration are
displayed in panels (d)--(f): bulk, normal and
parallel to the surface, respectively. The stability of the three different variants of
the pyramid is essentially identical, all three of them being unstable.}
\label{Fig_rpyramid}
\end{figure}

The next species of stationary lattice solutions is a pyramid-shaped
structure, with characteristics displayed in Fig.~\ref{Fig_rpyramid}, whose
base is a rhombus composed of four sites. The remaining out-of-plane vertex
site must have phase $0$ or $\pi $, since the phase values $\pi /2$ and 
$3\pi /2$ at this site do not produce a solution. The full set of pyramids
(bulk, normal, parallel ---see panels (d)--(f) of Fig.~\ref%
{Fig_rpyramid}) is \emph{completely unstable}, as seen in panel (c) of
Fig.~\ref{Fig_rpyramid}, the surface producing no stabilizing effect on it.
This strong instability actually arises at the lowest order in the
analytical eigenvalue calculations, which yield $\lambda =2\sqrt{5}%
\varepsilon i$, $\lambda =2\sqrt{2}\varepsilon i$, $\lambda =2\varepsilon $,
$\lambda =0$, and $\lambda =\mathcal{O}(\varepsilon ^{2})$, once again in
very good agreement with the full numerical results of Fig.~\ref%
{Fig_rpyramid}.

\subsection{Small-amplitude modes in a finite lattice\label{small-amp}}

Since our numerical investigation of the surface modes uses a finite
lattice, which allow the existence of small-amplitude modes (ones with the
zero threshold in terms of the norm --- cf.~discussion in Sec.~\ref%
{sec:TheorBack}), here we briefly consider the modes in a finite lattice
having the small-amplitude limit. Our aim is to show that these modes belong
to lower branches, as compared with the ``normal" surface modes considered
above. To this end, we concentrate on the lattice of size $M\!\times
M\!\times M$ lattice, subject to the zero boundary conditions, which imply
that discrete Laplacian (\ref{stat}) is modified at surfaces $n_{j}=1$ and $%
n_{j}=M$ ($j=1,...,3$) by setting the fields at sites $\mathbf{n}-\mathbf{e}%
_{j}$ and $\mathbf{n}+\mathbf{e}_{j}$, respectively, equal to zero. For the
sake of definiteness, we fix here $s=-1$ in Eq.~(\ref{stat}).

To determine the norm $N$ of small-amplitude modes we follow Ref.~\cite%
{BluKon}, and look for a solution to Eq.~(\ref{schr2}) with the amplitude $%
u_{\mathbf{n}}$ and coupling constant $\varepsilon $ being represented as
series 
\begin{eqnarray}
u_{\mathbf{n}} &=&\epsilon u_{0,\mathbf{n}}+\epsilon ^{2}u_{2,\mathbf{n}}+%
\mathcal{O}(\epsilon ^{3}),  \notag \\
&&  \label{eps} \\
\varepsilon &=&\varepsilon _{0}+\epsilon ^{2}\varepsilon _{2}+\mathcal{O}%
(\epsilon ^{3}),  \notag
\end{eqnarray}
in powers of small parameter $\epsilon\equiv\sqrt{8N/(M+1)^{3}}\ll 1$, which
vanishes in the limit of the infinite lattice ($M\rightarrow \infty $); in
other words, small $\epsilon $ characterizes the ``largeness" of the
lattice. We focus on real solutions here.

\begin{figure}[tbp]
\begin{center}
\includegraphics[width=\figw,angle=0]{\rootfig 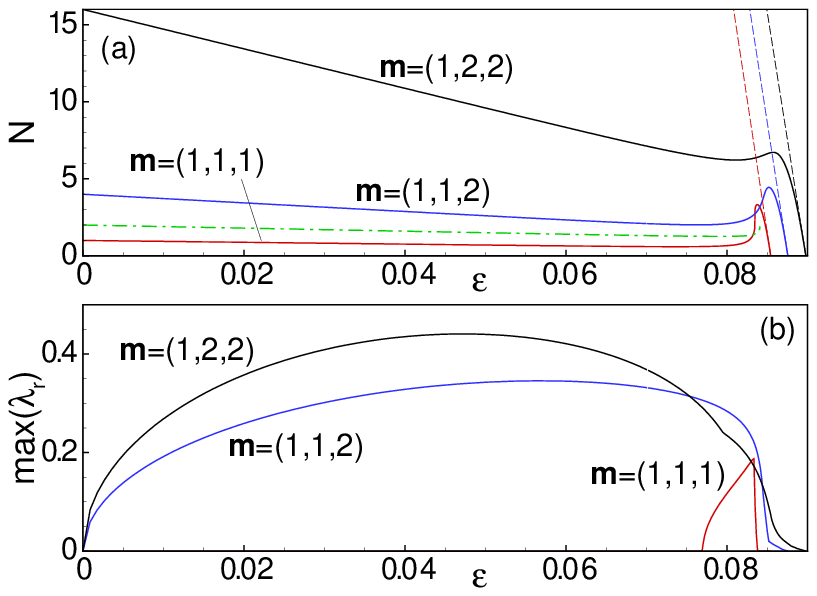}
\end{center}
\par
\vspace{-0.3cm}
\caption{(Color Online) Low-amplitude modes in a finite grid of size $%
9\!\times 9\!\times 9$ with $\Lambda =1.0$. (a) Norm $N$
versus coupling constant $\protect\varepsilon $ for several modes whose
low-amplitude limit is parameterized by vectors $\mathbf{m}$, as calculated
numerically and predicted by approximation (\protect\ref{eq:linlim}) (solid
and dashed lines, respectively). For comparison, dash-dotted line depicts
the norm for surface normal dipole. (b) Real part of
the critical (in)stability eigenvalue, calculated numerically. }
\label{fig:N-lam}
\end{figure}

Substituting expansions (\ref{eps}) into Eq.~(\ref{schr2}) and gathering
terms of the same order in $\epsilon $, we rewrite Eq.~(\ref{schr2}) in the
form of a set of equations:
\begin{equation}
\Lambda u_{j,\mathbf{n}}-\varepsilon _{0}\Delta u_{j,\mathbf{n}}=F_{j,%
\mathbf{n}}.  \label{eq:mul-scale}
\end{equation}
Here $F_{0,\mathbf{n}}=0$, $F_{2,\mathbf{n}}=\Lambda (\varepsilon
_{2}/\varepsilon _{0})u_{0,\mathbf{n}}+\left( u_{0,\mathbf{n}}\right) ^{3}$,
hence Eq.~(\ref{eq:mul-scale}) with $j=0$ gives rise to a linear eigenmode,
\begin{equation}
u_{0,\mathbf{n}}^{(\mathbf{m})}=\prod_{j=1}^{3}\sin \left( \frac{\pi
n_{j}m_{j}}{M+1}\right) ,  \label{eq:Q0}
\end{equation}
with the respective approximation for the lattice coupling constant,
\begin{equation}
\varepsilon _{0}^{(\mathbf{m})}=\Lambda \left[ 6+2\sum_{j=1}^{3}\cos \left(
\frac{\pi m_{j}}{M+1}\right) \right] ^{-1},
\end{equation}
parameterized by vector $\mathbf{m}=(m_{1},m_{2},m_{3})$. At the
same time, considering the solvability conditions for $j=2$, which
amounts to demanding the orthogonality of $F_{2,\mathbf{n}}$ and
$u_{0,\mathbf{n}}$, we obtain corrections to the coupling
constants,
\begin{equation}
\varepsilon =\varepsilon _{0}^{(\mathbf{m})}-\frac{\epsilon ^{2}\varepsilon
_{0}^{(\mathbf{m})}}{64\Lambda }\prod_{j=1}^{3}\left( 3+\delta
_{m_{j},(M+1)/2}\right) .  \label{eq:lam-small-ampl}
\end{equation}
It follows from Eq.~(\ref{eq:lam-small-ampl}) that each of the linear modes (%
\ref{eq:Q0}) is uniquely extended into a small-amplitude nonlinear one.
These modes are characterized by the linear dependence of the norm on
coupling constant $\varepsilon $:
\begin{equation}
N^{(\mathbf{m})}=\frac{8\Lambda (M+1)^{3}\left( \varepsilon _{0}^{(\mathbf{m}%
)}-\varepsilon \right) }{\varepsilon _{0}^{(\mathbf{m})}\prod_{j=1}^{3}%
\left( 3+\delta _{m_{j},(M+1)/2}\right) }.  \label{eq:linlim}
\end{equation}
From Eq.~(\ref{eq:linlim}) it follows that each mode,
parameterized by vector $\mathbf{m}$, exists when $\varepsilon $
belongs to the interval $0\leq \varepsilon \leq \varepsilon
_{0}^{(\mathbf{m})}$. The validity of approximation
(\ref{eq:linlim}) is corroborated by the coincidence of
analytical and numerical results in the vicinity of $\varepsilon _{0}^{(%
\mathbf{m})}$ (as shown in Fig.~\ref{fig:N-lam}), where these modes reaches
their small-amplitude limit. Such a property of these modes differs
considerably from the case of the surface modes which do not possess the
small-amplitude limit and require some minimal value of the norm (for the
normal dipole, depicted in Fig.~\ref{fig:N-lam} by dash-dotted line, the
minimal norm is $\approx 1.262$). Panel (b) in Fig.~\ref{fig:N-lam}
shows that only the mode, parameterized by vector $\mathbf{m}=(1,1,1)$, is
stable for $\varepsilon $ close to $\varepsilon _{0}^{(\mathbf{m})}$, while
other modes are completely unstable.

\section{Dynamics}

\begin{figure}[tbp]
\begin{tabular}{|c|}
\hline
\includegraphics[width=8cm,height=4.75cm,angle=0]{\rootfig
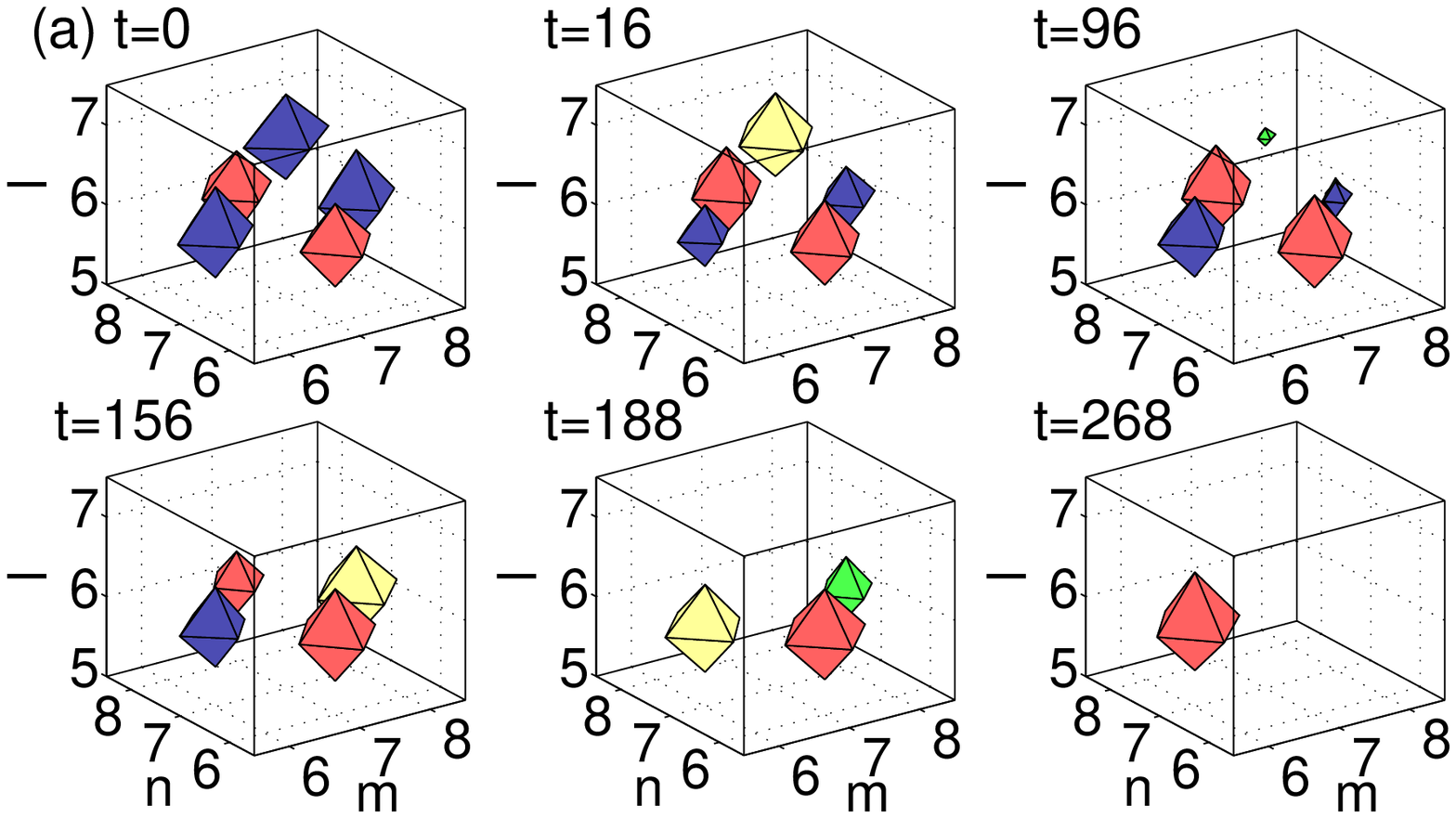} \\ \hline
\includegraphics[width=8cm,height=4.75cm,angle=0]{\rootfig
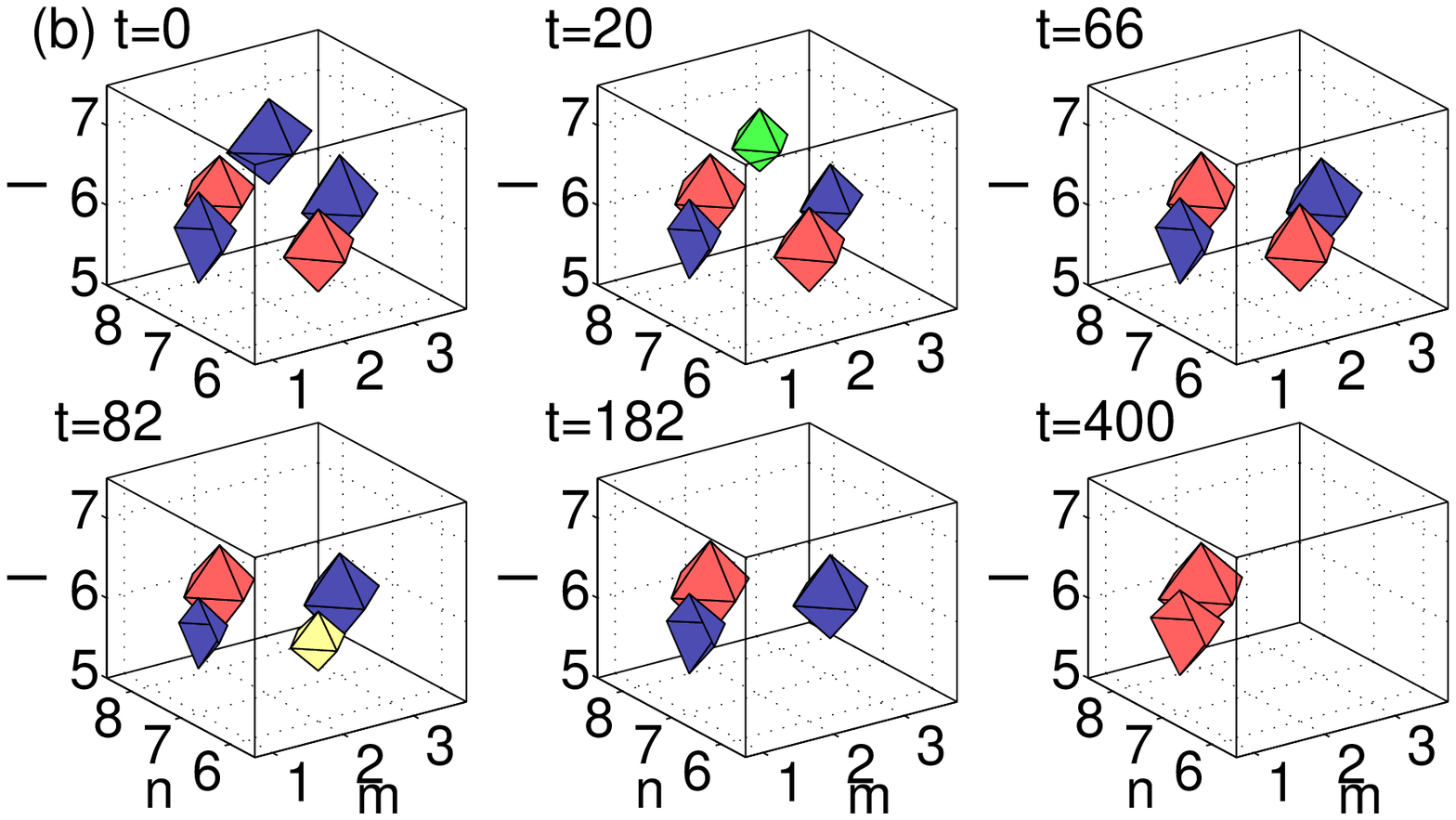} \\ \hline
\includegraphics[width=8cm,height=4.75cm,angle=0]{\rootfig
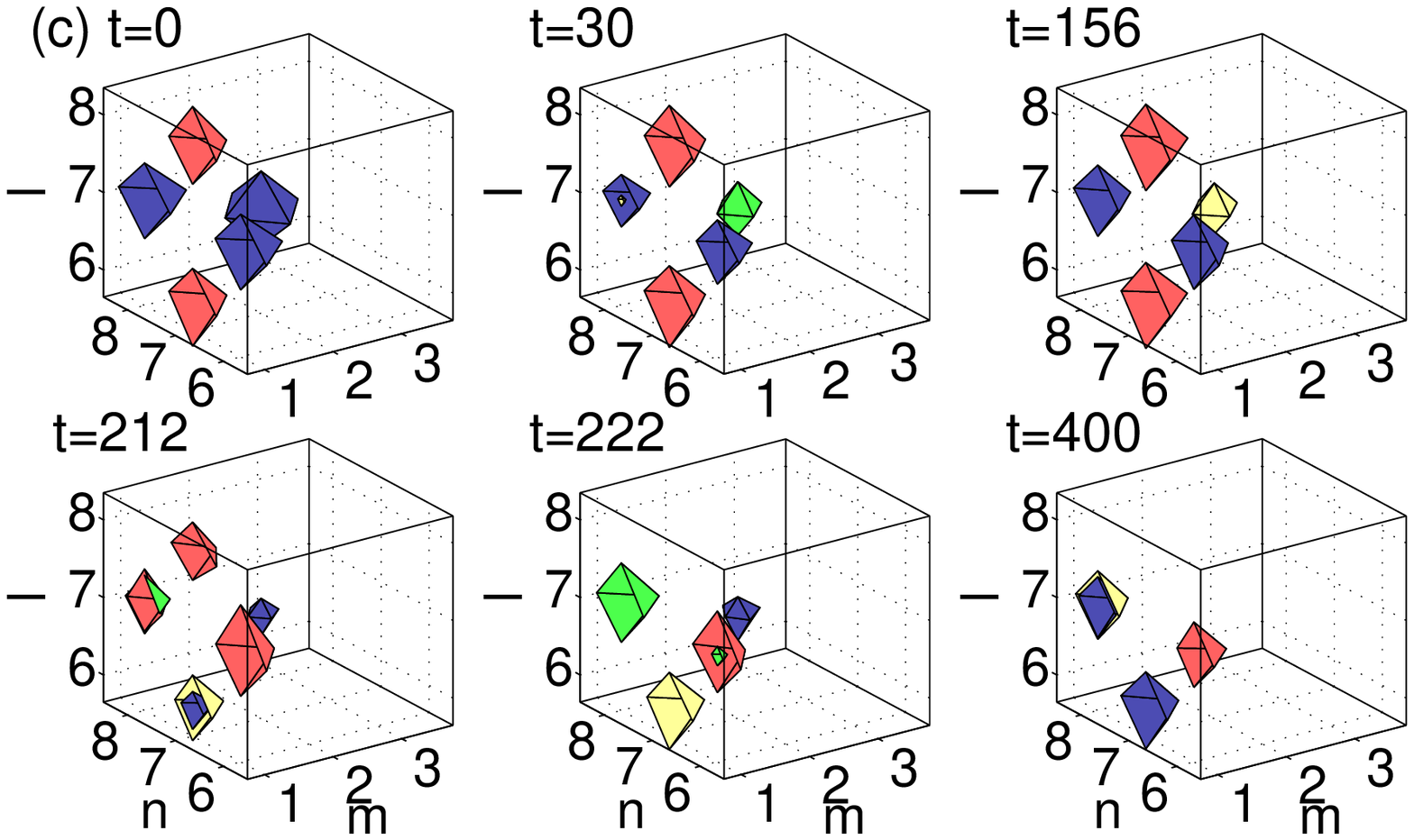} \\ \hline
\end{tabular}%
\caption{(Color Online) The evolution of unstable pyramids. Panels (a), (b),
and (c) display, respectively, the transformation of a bulk pyramid, and of
ones oriented normally and parallel to the surface, for $\protect\varepsilon %
=0.2$. }
\label{fig_dyn_pyr}
\end{figure}

In this section we examine the nonlinear evolution of the various
configurations, displaying the results in a set of figures (see Figs.~\ref%
{fig_dip}--\ref{fig_dyn_pyr}). In each case, the evolution is initiated at a
value of the coupling $\varepsilon $ taken \emph{beyond} the instability
threshold, and an initial small random perturbation is applied in order to
expedite the onset of the instability.

\begin{figure}[tbp]
\begin{tabular}{|c|}
\hline
\includegraphics[width=8cm,angle=0]{\rootfig 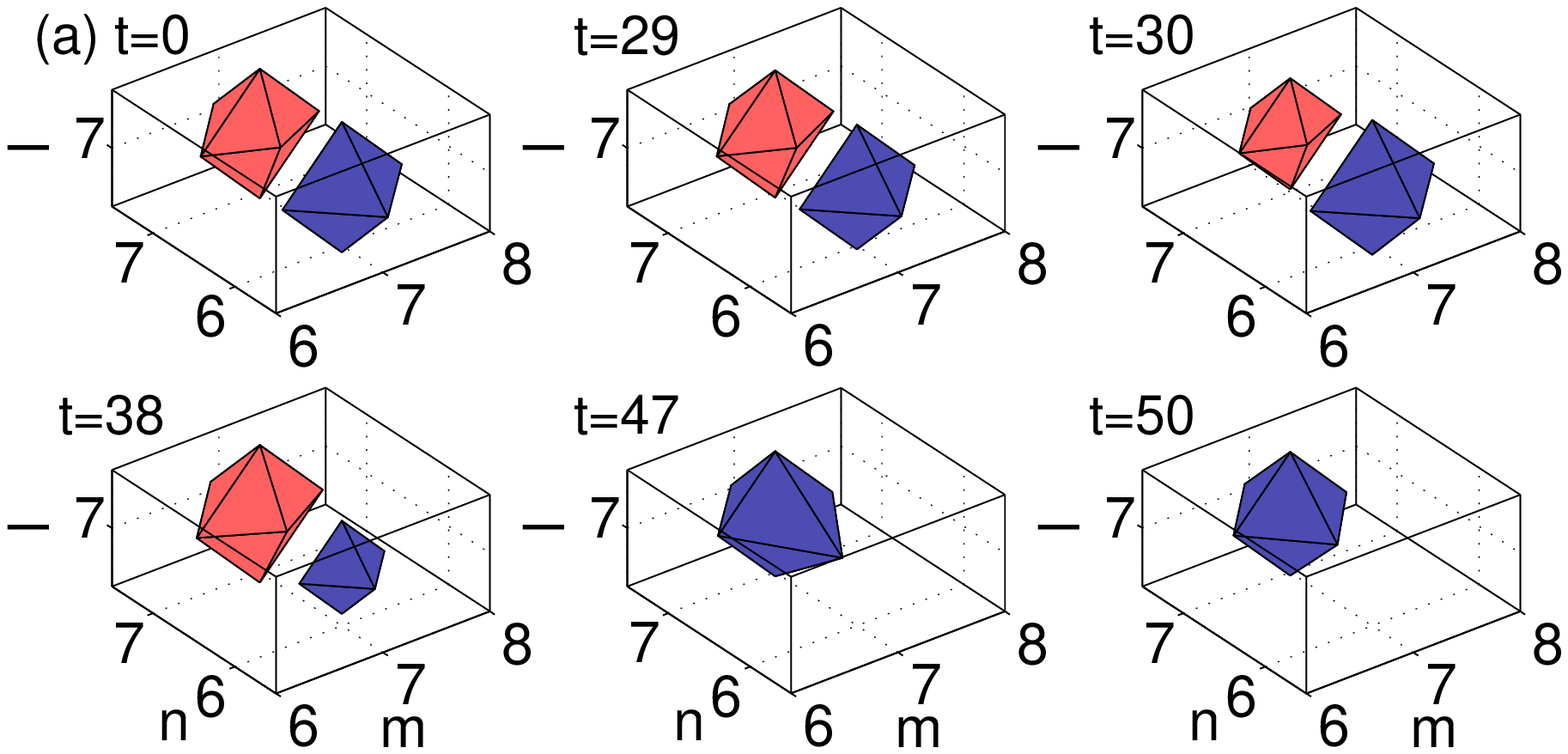} \\ \hline
\includegraphics[width=8cm,angle=0]{\rootfig 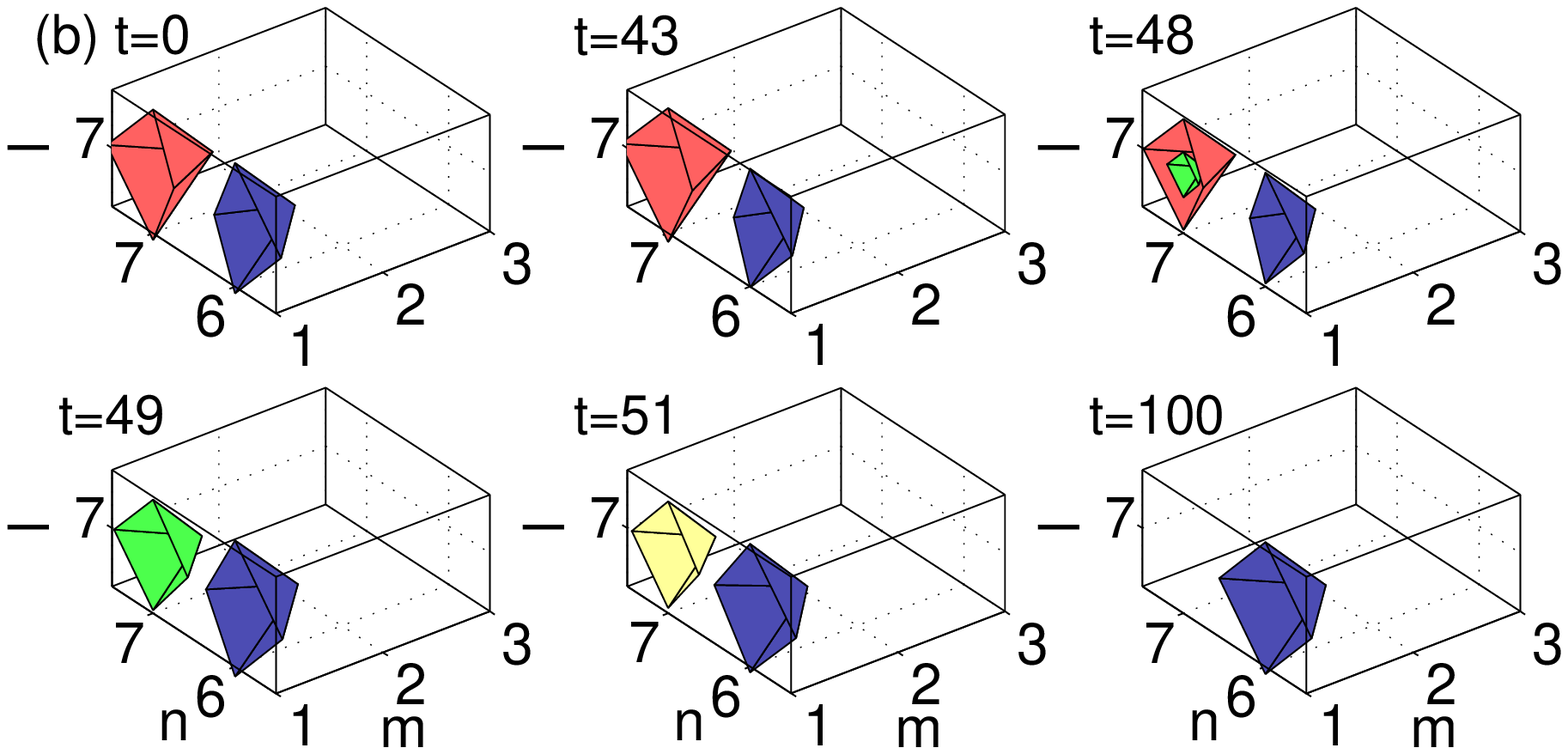} \\ \hline
\includegraphics[width=8cm,angle=0]{\rootfig 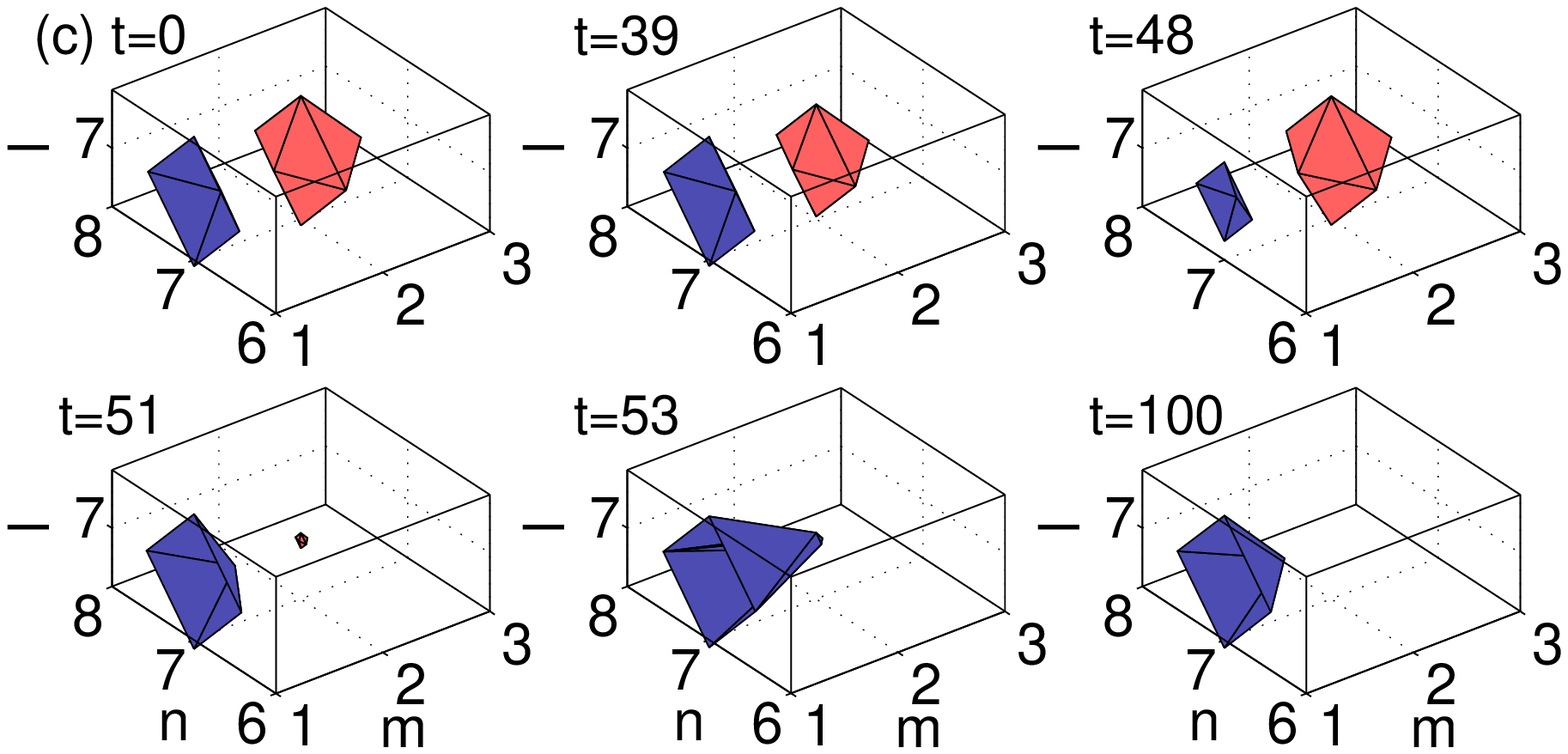} \\ \hline
\end{tabular}%
\caption{(Color Online) The evolution of unstable dipoles: (a) a bulk dipole;
(b) and (c) dipoles placed parallel and normally to the surface, respectively.
In all the cases, the dipole is subject to oscillatory instability, which is
responsible for the eventual concentration of most of the norm at a single
site (i.e., the transition to a monopole). Parameters are $\Lambda =1$, $%
\protect\varepsilon =0.2$, the lattice has a size of $13\!\times 13\!\times
13$, and times are indicated in the panels. All iso-contour plots are
defined as $\mathrm{Re}(u_{l,n,m})=\pm 0.75=\mathrm{Im}(u_{l,n,m})$, and the
initial configurations were perturbed with random noise of amplitude $0.01$.
The coding for the iso-contours is as follows: dark gray (blue) and gray
(red) colors pertain to iso-contours of the real part of the solutions,
while the light gray (green) and very light gray (yellow) colors correspond
to the iso-contours of the imaginary part. }
\label{fig_dip}
\end{figure}

\begin{figure}[tbp]
\begin{tabular}{|c|}
\hline
\includegraphics[width=8cm,height=4.35cm,angle=0]{\rootfig
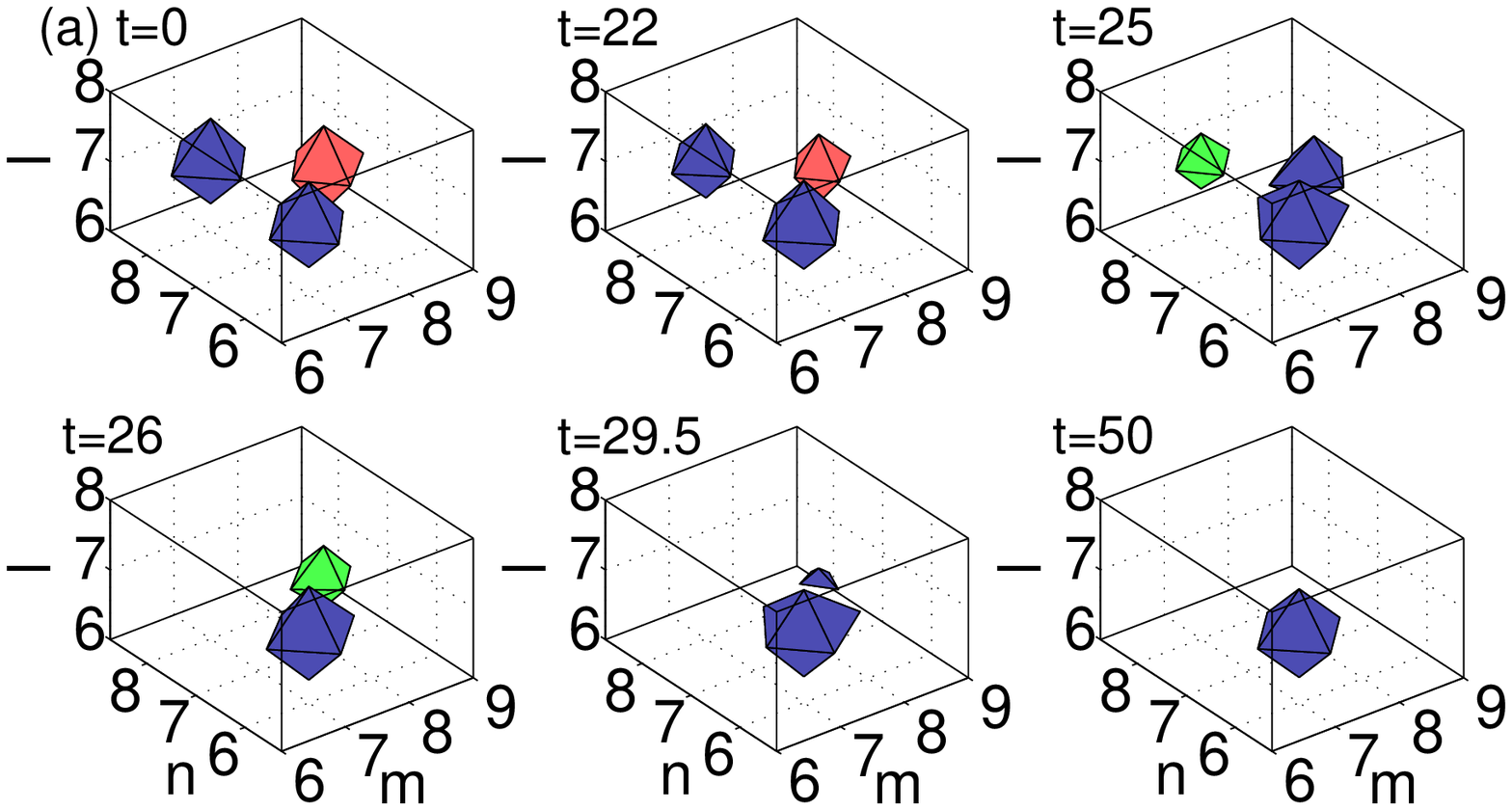} \\ \hline
\includegraphics[width=8cm,height=4.35cm,angle=0]{\rootfig
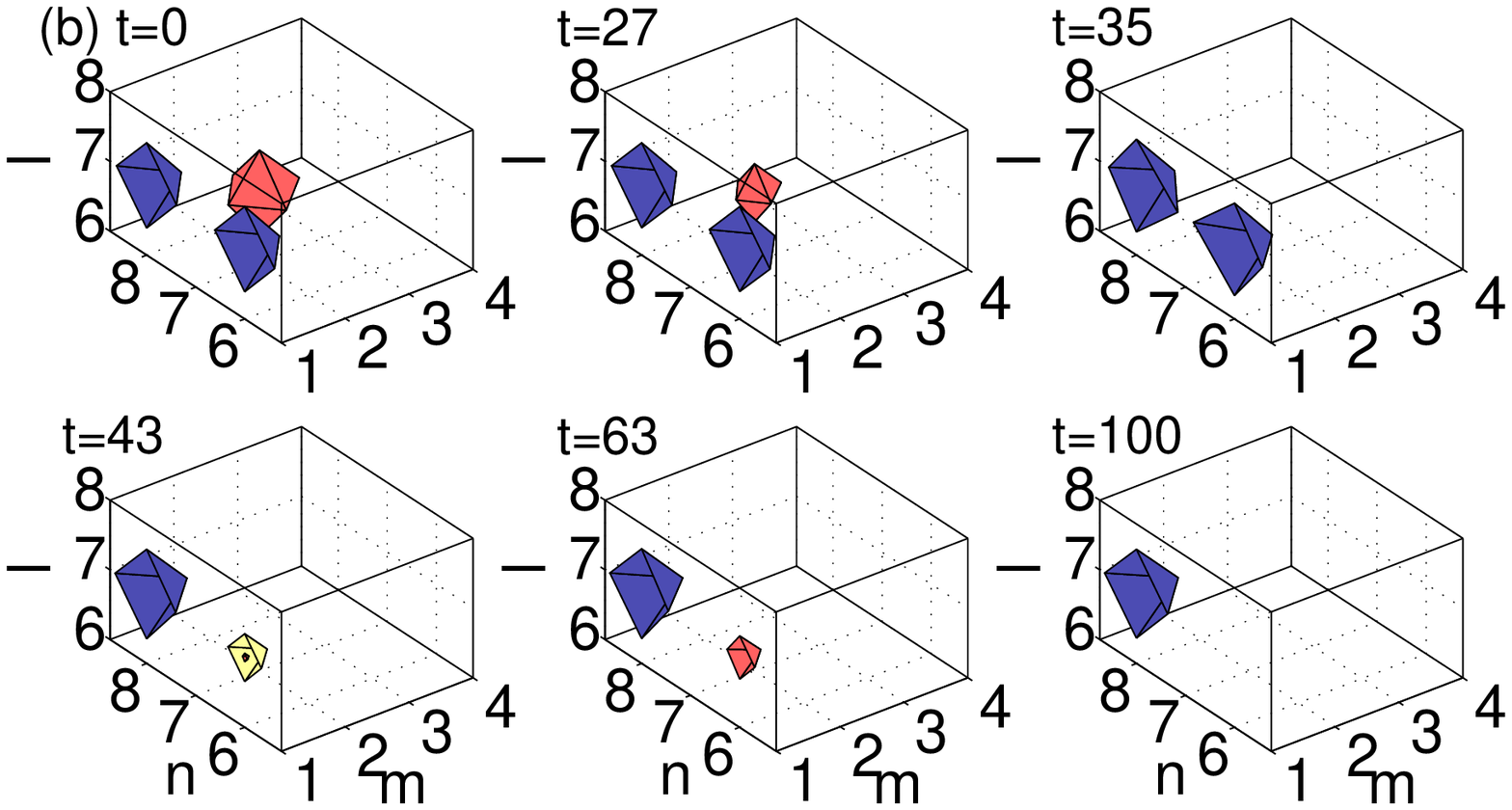} \\ \hline
\end{tabular}%
\caption{(Color Online) The evolution of the unstable three-site horseshoes:
(a) bulk three-site horseshoe and (b) the horseshoe oriented normally to the
surface. In both cases, the unstable horseshoe is subject to an oscillatory
instability, which leads to the eventual concentration of most of the norm
in a single-site structure. The iso-contours and parameters are the same as
in Fig.~\protect\ref{fig_dip} except that $\protect\varepsilon =0.3$. }
\label{fig_3horse}
\end{figure}

\begin{figure}[tbp]
\begin{tabular}{|c|}
\hline
\includegraphics[width=8cm,height=4.20cm,angle=0]{\rootfig
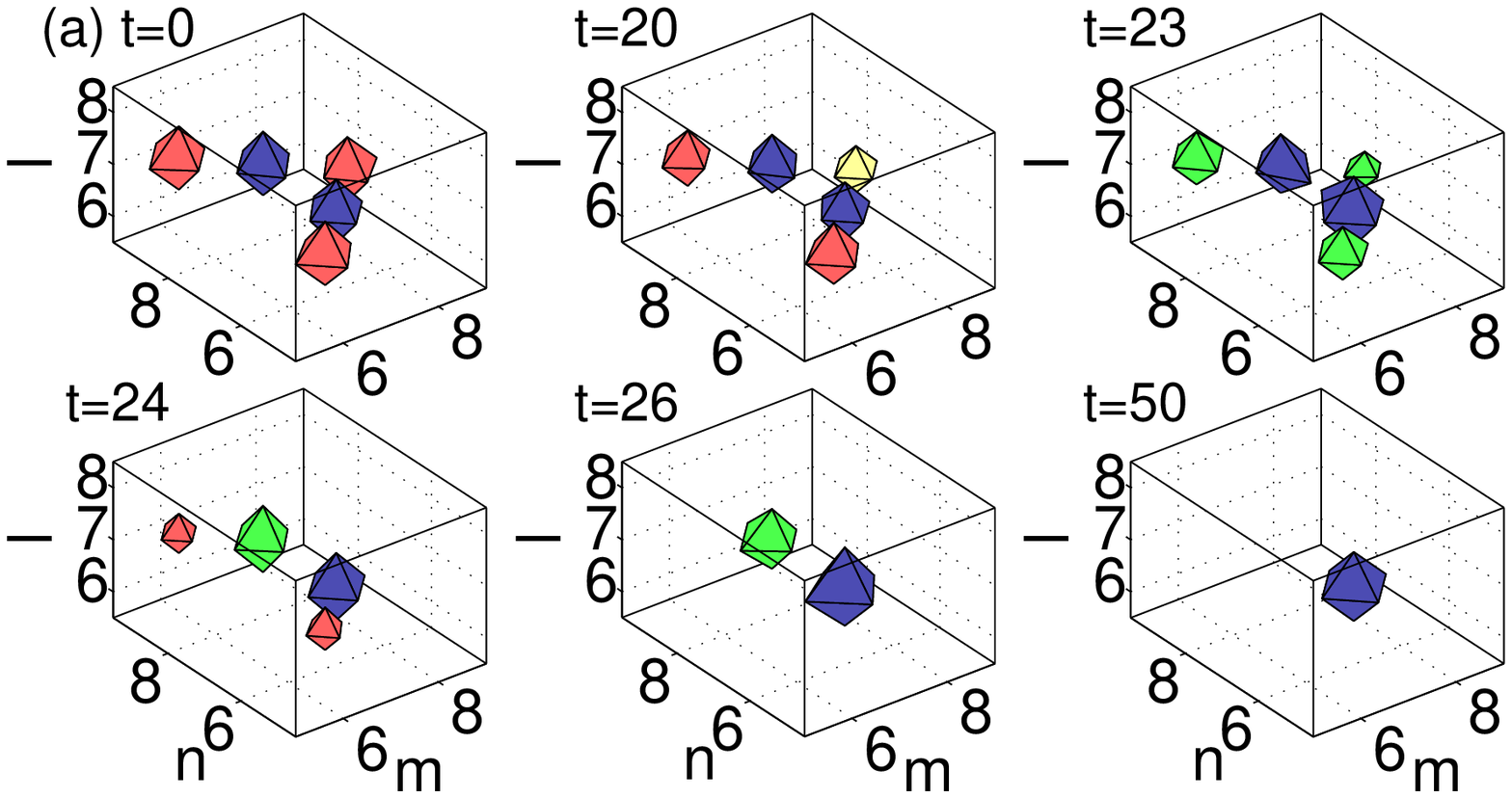} \\ \hline
\includegraphics[width=8cm,height=4.20cm,angle=0]{\rootfig
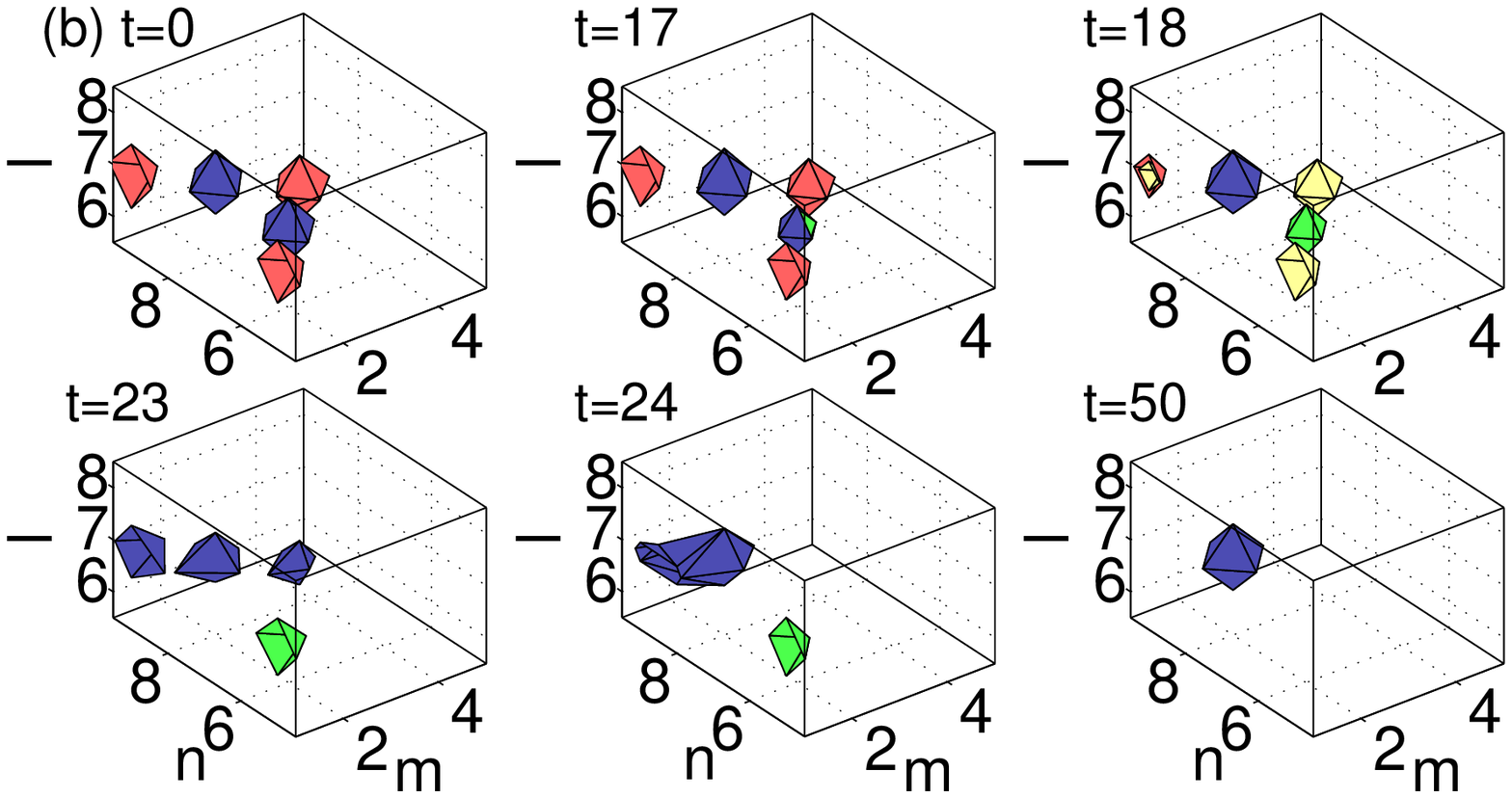} \\ \hline
\end{tabular}%
\caption{(Color Online) The evolution of unstable five-site horseshoes: (a)
the bulk horseshoe, and (b) the five-site horseshoe oriented normally to the
surface. In both cases, the unstable horseshoe is subject to an oscillatory
instability, which triggers the transition to a monopole. The iso-contours
and parameters are the same as in Fig.~\protect\ref{fig_3horse}. }
\label{fig_5horse}
\end{figure}

\begin{figure}[tbp]
\begin{tabular}{|c|}
\hline
\includegraphics[width=8cm,height=4.75cm,angle=0]{\rootfig
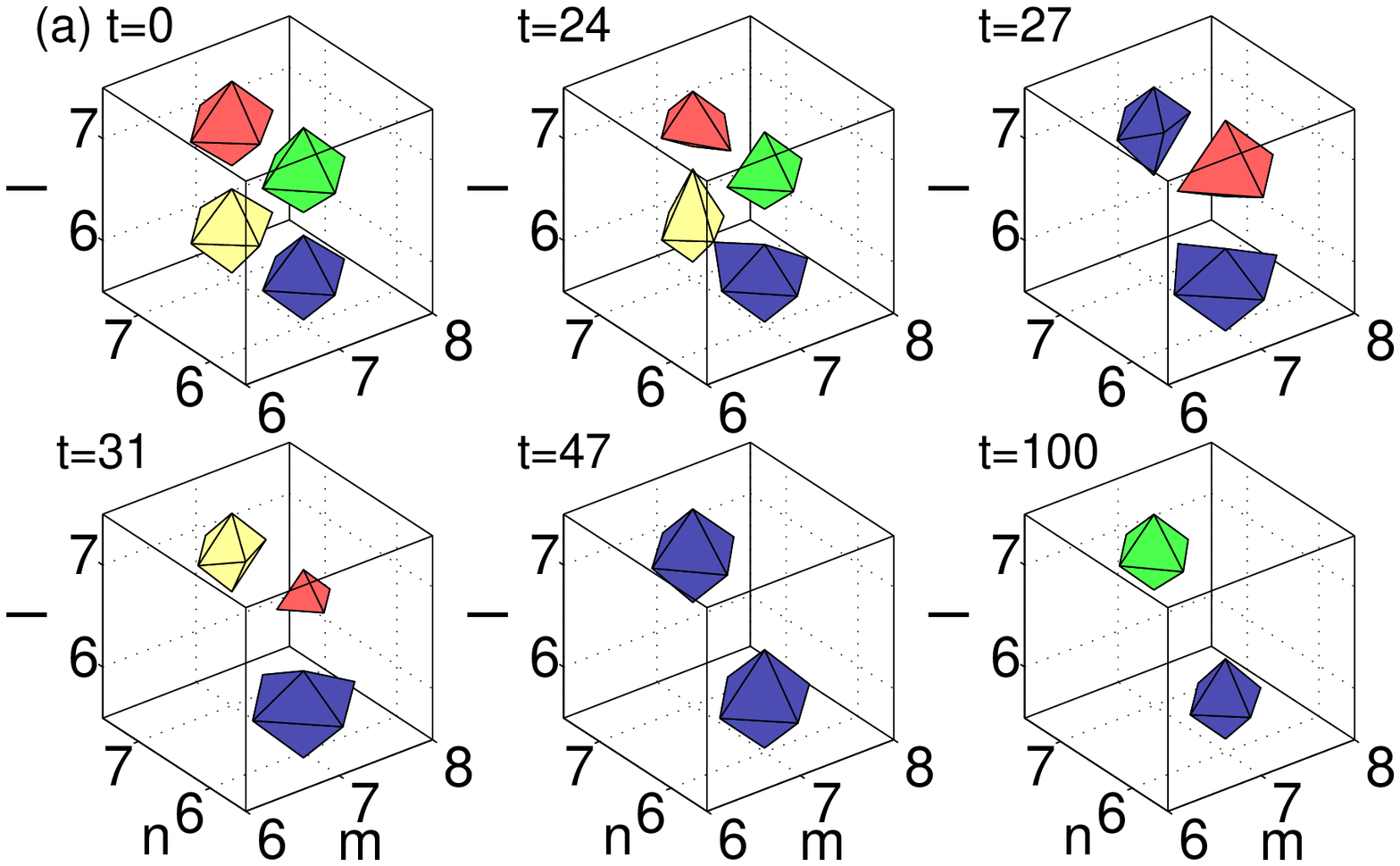} \\ \hline
\includegraphics[width=8cm,height=4.75cm,angle=0]{\rootfig
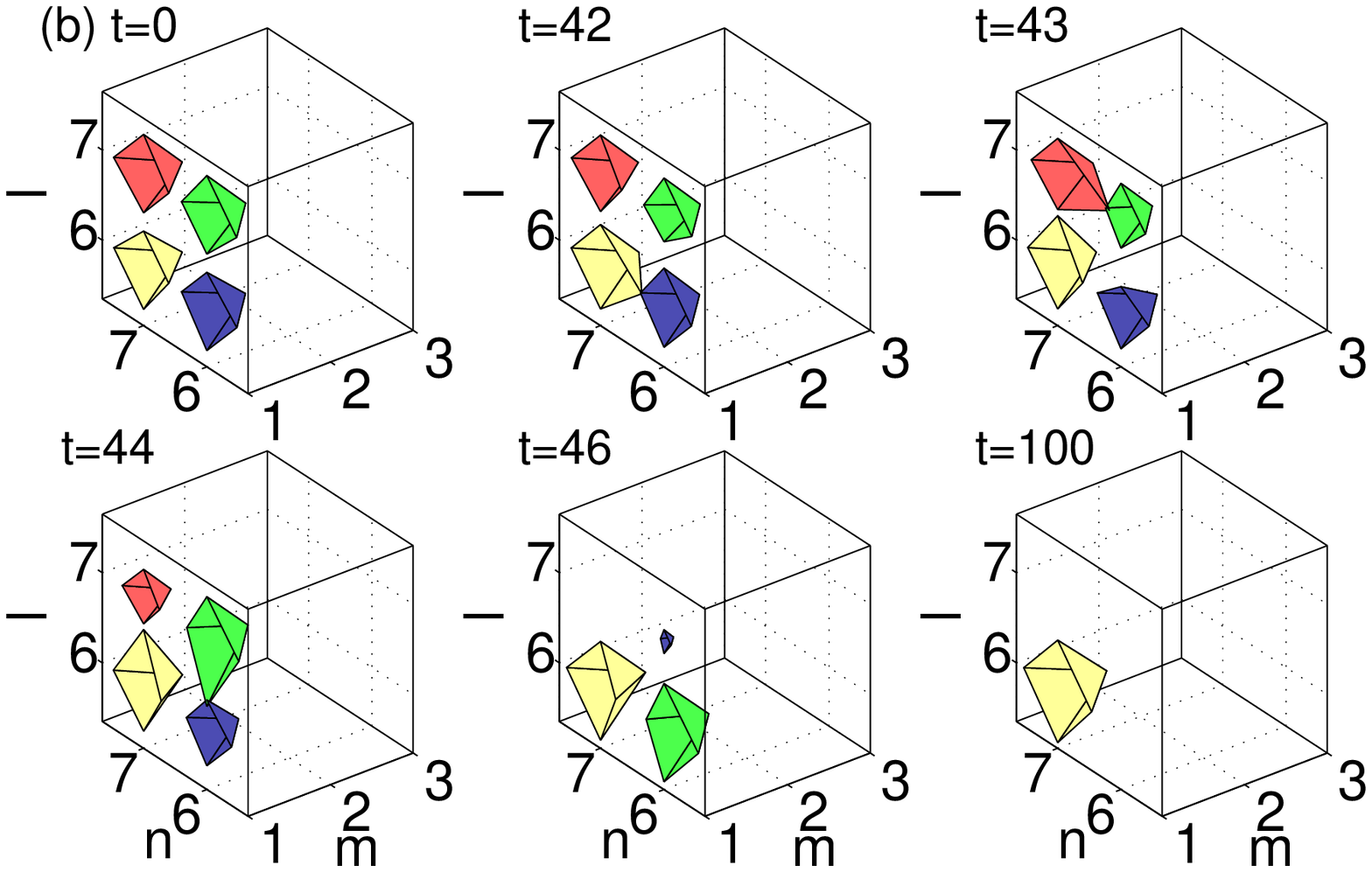} \\ \hline
\end{tabular}%
\caption{(Color Online) The evolution of unstable vortices: (a) the bulk
vortex for $\protect\varepsilon =0.3$ and (b) the vortex parallel to the
surface, for $\protect\varepsilon =0.6$ and $\Lambda =1$. The iso-contour
plots are defined by $\mathrm{Re}(u_{l,n,m})=\pm 1=\mathrm{Im}(u_{l,n,m})$. }
\label{fig_vortex}
\end{figure}

All the figures display the evolution of the instability at six different
moments of time, starting at $t=0$, and ending at a time well beyond the
point at which the instability manifests itself. All configurations that
were predicted above to be unstable through nonzero real parts of the
(in)stability eigenvalue $\lambda $ indeed exhibit instability dynamics,
which eventually results in a transition to a different configuration. In
the case of the dipoles and horseshoes, Figs.~\ref{fig_dip}--\ref{fig_5horse}
show a spontaneous transition to monopole patterns, i.e., ones centered
around a single excited site. On the other hand, in the case of the vortices
and pyramids shown in Figs.~\ref{fig_vortex}--\ref{fig_dyn_pyr}, a few sites
may remain essentially excited at the end of the evolution. The monopole is,
obviously, the most robust dynamical state in the lattice system, with the
widest stability interval, in comparison with other discrete structures.
This simplest state becomes unstable, for given $\Lambda $, only at values
of the coupling constant $\varepsilon \approx \Lambda $ \cite{earlier3}.
Another structure with a relatively wide stability region is the dipole (the
more stable the wider the distance between its constituent sites \cite%
{vor3dnew}), consonant with the observation that some of the structures
(especially ones with a large number of excited sites, such as vortices and
pyramids) dynamically transform into dipoles.

Generally speaking, the exact scenario of the nonlinear evolution and the
finally established state depend on details of the initial perturbation. In
the figures, each configuration is shown by iso-level contours of distinct
hues. In particular, dark gray (blue) and gray (red)
are iso-contours of the real part of the solutions, while the light gray
(green) and very light gray (yellow) colors
depict the imaginary part of the same solutions.

A case that needs further consideration is that of the three-site horseshoe.
As observed from the stability analysis presented in Fig.~\ref{Fig_3_horse},
this horseshoe in the bulk gives rise to a small unstable purely real
eigenvalue for all values of $\varepsilon $, see the lower green
dashed-dotted curve in panel (c) of the figure. Despite the presence
of this eigenvalue, the evolution of the unstable bulk three-site horseshoe
is predominantly driven by the unstable complex eigenvalues, if any (in
fact, for $\varepsilon >0.226$, see the dashed-dotted (green) line 
of Fig.~\ref{Fig_3_horse}.(c)). A careful analysis of the
instability corresponding to the small purely real eigenvalue for $%
\varepsilon <0.226$ (i.e., before the complex eigenvalues become unstable)
reveals that the corresponding dynamics amounts to an extremely weak
exchange of the norm between the two in-phase excited sites (see Fig.~\ref%
{Fig_3_horse}). The norm exchange is driven by the corresponding unstable
eigenfunction, which looks like a dipole positioned at the two
aforementioned in-phase sites. The difficulty in observing this evolution
mode is explained by the fact that, in the course of the norm exchange, only
$\sim 0.1\%$ of the total norm is actually transferred between the two
sites. 
Furthermore, as mentioned earlier, the corresponding small real eigenvalue
is completely suppressed by the surface (see panel (c) in Fig.~\ref%
{Fig_3_horse}). It is worth noting that such stable three-site horseshoe
surface structures may also be generated by the evolution of more complex
unstable waveforms, such as the five-site pyramids placed normally to the
surface, see the bottom panel in Fig.~\ref{fig_dyn_pyr}.


\section{Conclusions}

In this work, we have investigated localized modes in the vicinity of a
two-dimensional surface, in the framework of the three-dimensional DNLS
equation, which is a prototypical model of nonlinear dynamical lattices. We
have found that the surface may readily stabilize localized structures that
are unstable in the bulk (such as three-site horseshoes), and, on the other
hand, it may inhibit the formation of some other structures that exist in
the bulk (such as vortices which are oriented normally to the surface,
although ones parallel to the surface do exist and have their stability
region; 
a qualitative explanation to these features was proposed, based on the
analysis of the interaction of the vortical state with its ``mirror image"). 
The most typical surface-induced effect is the expansion of the stability
intervals for various solutions that exist in the bulk and survive in the
presence of the surface. This feature may be attributed to the decrease,
near the surface, of the number of neighbors to which excited sites couple,
since the approach to the continuum limit, i.e., the strengthening of the
linear couplings to the nearest neighbors, is responsible for the onset of
the instability or disappearance of all the localized stationary states in
the three-dimensional dynamical lattice.

On the other hand, while the techniques elaborated in Refs.~\cite%
{Pelinovsky1d,Pelinovsky,Pelinovsky3d} for the analysis of localized states
in bulk lattices are quite useful in understanding the dominant stability
properties of the solutions, the surface gives rise to specific effects,
such as the stabilization of higher-order solutions or the suppression of
some types of vortex structures, which cannot be explained by these methods.
Therefore, it would be very relevant to modify these techniques, which are
based on the Lyapunov-Schmidt reductions, so as to take the presence of the
surface into regard.

\end{document}